\documentclass[symmetry, review, accept, pdftex, moreauthors]{Definitions/mdpi} 

\firstpage{1} 
\pubvolume{1}
\issuenum{1}
\articlenumber{1}
\pubyear{2025}
\copyrightyear{2025}
\datereceived{ } 
\daterevised{ } 
\dateaccepted{ } 
\datepublished{ } 
\hreflink{https://doi.org/} 
\DeclareMathAlphabet{\mathpzc}{OT1}{pzc}{m}{it}

\Title{\texorpdfstring{Electroexcitation of Nucleon Resonances\\ and the Emergence of Hadron Mass}{Electroexcitation of Nucleon Resonances and the Emergence of Hadron Mass}}

\TitleCitation{Electroexcitation of Nucleon Resonances and the Emergence of Hadron Mass}

\Author{Patrick Achenbach $^{1}$\orcidE{}, 
Daniel S.\ Carman $^{1}$\orcidA{}, 
Ralf  W.\ Gothe $^{2}$\orcidC{},\\
Kyungseon Joo $^{3}$\orcidK{},
Victor I.\ Mokeev $^{1}$\orcidB{}, 
and Craig D.\ Roberts $^{4,5}$\orcidD{}\,*}

\AuthorNames{Patrick Achenbach, Daniel S. Carman, Ralf W. Gothe, Kyungseon Joo, Victor I. Mokeev, and Craig D. Roberts}

\AuthorCitation{Achenbach, P.; Carman, D.S.; Gothe, R.W.; Joo, K.; Mokeev, V.I.; and Roberts, C.D.}

\address{%
$^{1}$ \quad  \href{https://ror.org/02vwzrd76} {Thomas Jefferson National Accelerator Facility}, Newport News, Virginia 23606, USA\\
$^{2}$ \quad \href{https://ror.org/02b6qw903}{University of South Carolina}, Columbia, South Carolina 29208, USA \\
$^{3}$ \quad \href{https://ror.org/02der9h97}{University of Connecticut}, Storrs, Connecticut 06269, USA \\
$^{4}$ \quad School of Physics, \href{https://ror.org/01rxvg760}{Nanjing University}, Nanjing, Jiangsu 210093, China \\
$^{5}$ \quad Institute for Nonperturbative Physics, \href{https://ror.org/01rxvg760}{Nanjing University}, Nanjing, Jiangsu 210093, China \\
}

\corres{Correspondence:
\href{patricka@jlab.org}{patricka@jlab.org} (PA); 
\href{carman@jlab.org}{carman@jlab.org} (DSC); 
\href{mailto:gothe@usc.edu}{gothe@sc.edu} (RWG); 
\href{mailto:kyungseon.joo@uconn.edu}{kyungseon.joo@uconn.edu} (KJ); 
\href{mokeev@jlab.org}{mokeev@jlab.org} (VIM); 
\href{mailto:cdroberts@nju.edu.cn}{cdroberts@nju.edu.cn} (CDR)
 \\[2ex]
\hspace*{26em}\sf\emph{Preprint no}.\ NJU-INP 102/25, JLAB-PHY-25-4339
}

\abstract{
Developing an understanding of phenomena driven by the emergence of hadron mass (EHM) is one of the most challenging problems in the Standard Model. 
This discussion focuses on the impact of results on nucleon resonance \texorpdfstring{($N^\ast$)}{} electroexcitation amplitudes (or \texorpdfstring{$\gamma_vpN^\ast$}{} electrocouplings) obtained from experiments during the 6-GeV era in Hall~B at Jefferson Lab on understanding EHM. 
Analyzed using continuum Schwinger function methods (CSMs), these results have revealed new pathways for the elucidation of EHM. 
A good description of the \texorpdfstring{$\Delta(1232)3/2^+$}{Delta(1232)+}, \texorpdfstring{$N(1440)1/2^+$}{N(1440)1/2+}, and \texorpdfstring{$\Delta(1600)3/2^+$}{N(1600)3/2+} electrocouplings, achieved by CSM analyses that express a realistic dressed quark mass function, 
sheds light on the strong interaction dynamics that underlies EHM. 
Extensions to \texorpdfstring{$N^\ast$}{nucleon resonance} studies for higher-mass states are outlined, as well as experimental results anticipated in the 12-GeV era at Jefferson Lab and those that would be enabled by a further increase of the beam energy to 22~GeV.}

\keyword{Nucleon resonances; exclusive meson electroproduction; hadron structure; electromagnetic probes; emergence of hadron mass; continuum Schwinger function methods; quantum chromodynamics
}

\begin{document}

\section{Introduction}

Understanding the emergence of the dominant component ($>$ 98\%) of visible mass in the Universe remains one of the most profound and unresolved challenges within the Standard Model (SM)~\cite{Leader:1996hm, Weinberg:1996kr} that describes the known fundamental interactions in Nature with the exception of gravity. 
Unlike the electroweak sector, where perturbative methods are sufficient to describe the current status of experimental investigations, the strong interaction undergoes a transition from the perturbative to strongly coupled regimes as probe energy scales are reduced.
In this process, it generates effective degrees of freedom, commonly referred to as dressed gluons and quarks, which seem well suited to describing hadron structure. 
Like their partonic seeds~\cite{Fritzsch:1972jv}, these entities are only observed indirectly in hadron structure experiments, yet they provide a physical basis that readily delivers a practicable understanding of real-world hadrons. 
The effective interactions between dressed quarks and gluons are nevertheless highly intricate and require careful theoretical treatment. 

The dressed gluons and quarks and their effective interactions emerge from the quantum chromodynamics (QCD) Lagrangian. 
However, their connection to the elementary structural objects in QCD, the partonic quarks and gauge gluons, and to their interactions, encoded in the QCD Lagrangian, which are relevant in perturbative QCD (pQCD), is highly non-trivial~\cite{Roberts:2015lja, Horn:2016rip, Roberts:2021nhw, Binosi:2022djx, Ding:2022ows, Carman:2023zke, Ferreira:2023fva, Raya:2024ejx}. 
Additionally, the aforementioned quasiparticle hadron constituents and the interactions between them exhibit significant evolution with length scales. 

These complexities suggest that the generation of both hadron mass and structure should be viewed as emergent phenomena. 
This is especially true given that the emergence of hadron mass (EHM) and the intricate structure of hadrons cannot be calculated from the QCD Lagrangian within any sort of perturbative expansion.
Perturbative QCD is unconnected with such phenomena because the formal power series expansion of any $S$-matrix element in all quantum field theories of interest is at best an asymptotic series~\cite{Dyson:1952tj}.
Instead, expanding our understanding of EHM and hadron structure requires a concerted, synergistic effort.
It must combine experimental research, phenomenological studies, and significant improvements in the description of the transition between the pQCD domain, whereupon perturbative expansions can deliver an approximation to some observables, and the strongly coupled (sQCD) regime, where nonperturbative treatments of the fundamental interactions encoded in the QCD Lagrangian are essential for any predictions~\cite{Roberts:2021nhw, Binosi:2022djx, Ding:2022ows, Carman:2023zke, Ferreira:2023fva, Raya:2024ejx}.

In recent decades, significant progress has been made in exploring the structure of nucleon excited states ($N^\ast$) using data on exclusive meson electroproduction off nucleons, primarily at Jefferson Lab (JLab)~\cite{Mokeev:2022xfo, Mokeev:2023zhq, Burkert:2022ioj, Aznauryan:2011qj, Carman:2016hlp}, along with results from MAMI~\cite{Mart:2014eoa, Sparveris:2013ena, A1:2017fvu, Blomberg:2019caf} and MIT Bates~\cite{MIT-BatesOOPS:2003afe}. 
The combination of the continuous electron beam from the CEBAF accelerator, with energies up to 6~GeV, and the nearly $4\pi$ acceptance CLAS detector in Hall~B at JLab~\cite{CLAS:2003umf}, has provided the majority of the worldwide data on most exclusive meson electroproduction channels off nucleons in the resonance region. 
Analyses of these results within well-constrained reaction models have enabled  extraction of the electroexcitation amplitudes, also called $\gamma_vpN^\ast$ electrocouplings, for $N^\ast$ states with masses ranging up to 1.75\,GeV and for virtual photon four-momentum transfer squared ($Q^2$), commonly referred to as photon virtuality, covering the domain $Q^2 \in [0, 5]$\,GeV$^2$~\cite{Aznauryan:2011qj, Proceedings:2020fyd, Mokeev:2022xfo, Burkert:2017djo}.

Herein, we discuss the impact of experimental $\gamma_vpN^\ast$ electrocoupling results on our understanding of the strong interaction dynamics that underlie EHM. 
The available experimental data on the $Q^2$ evolution of $\gamma_vpN^\ast$ electrocouplings for nucleon resonances with different structure, analyzed using continuum Schwinger function methods (CSMs), offer promising new insights into EHM. 
We will outline the prospects for continuing progress from these studies with data obtained during the ongoing 12-GeV era experiments using the CLAS12 detector in Hall~B at JLab~\cite{Burkert:2020akg}. 
Additionally, we will highlight the unique opportunity to explore the full domain of distances whereupon the dominant part of baryon masses and $N^\ast$ structure emerges in the transition from the pQCD to the sQCD regimes. 
The potential of such studies would be greatly enhanced by an increase of the JLab accelerator energy up to 22\,GeV~\cite{Accardi:2023chb}.

\section{EHM Basics}

Hadron mass and structure are generated by strong interactions; indeed, they are definitive of strong interactions. 
In the SM, the short-distance behavior of strong interactions is derived from the QCD Lagrangian, expressed in terms of partonic quarks and gauge gluons, which are the degrees of freedom used for pQCD computations, \textit{viz}.,
\begin{subequations}
\label{QCDdefine}
\begin{align}
{\mathpzc L}_{\rm QCD} & = \sum_{{\mathpzc f}=u,d,s,\ldots}\hspace{-1em}
\bar{q}_{\mathpzc f} \left[\gamma\cdot\partial
    + i g \tfrac{1}{2} \lambda^a\gamma\cdot A^a+ m_{\mathpzc f}\right] q_{\mathpzc f}
    + \tfrac{1}{4} G^a_{\mu\nu} G^a_{\mu\nu},\\
%
%
\label{gluonSI}
G^a_{\mu\nu} & = \partial_\mu A^a_\nu + \partial_\nu A^a_\mu -
g f^{abc}A^b_\mu A^c_\nu,
\end{align}
\end{subequations}
where $\{q_{\mathpzc f}\,|\,{\mathpzc f}=u,d,s,c,b,t\}$ are the fields associated with the six known flavors of quarks; 
$\{m_{\mathpzc f}\}$ are their current masses, generated by Higgs boson couplings; 
$\{A_\mu^a\,|\,a=1,\ldots,8\}$ are the color-octet gluon fields, whose matrix structure is encoded in $\{\tfrac{1}{2}\lambda^a\}$, the generators of SU$(3)$ color in the fundamental representation; 
and $g$ is the \emph{unique} QCD coupling, in terms of which one conventionally defines $\alpha = g^2/[4\pi]$. 
Much or even most of that which is nontrivial about QCD can be attributed to its being a Poincar\'e-invariant quantum non-Abelian gauge field theory defined in four-dimensional (4D) spacetime~\cite{Roberts:2022rxm}. 

The Lagrangian in Eq.\,(\ref{QCDdefine}) is expressed using a Euclidean metric, which will be employed throughout hereafter; see, \textit{e.g}., Ref.\,\cite[Sec.\,2.3]{Roberts:1994dr}. 
The choice of a Euclidean metric is practically essential for comparing CSM predictions with results obtained from the numerical simulation of lattice-regularized QCD (lQCD) because the effective use of lQCD requires a positive definite probability measure, which is only available in Euclidean space.
Moreover, there are good mathematical reasons to view Euclidean space as primary, since if there is any hope of arriving at a rigorous definition of QCD, then it is by formulating the theory in Euclidean space; see, \textit{e.g}., Ref.\,\cite[Sec.\,1]{Ding:2022ows}. 

For hadrons composed of light $u$- and $d$-quarks, the masses in Eq.\,\eqref{QCDdefine} are significantly smaller than the hadron masses. The limit where the Lagrangian current quark masses are set to zero defines the QCD chiral limit.  
A rigorous definition of the chiral limit is only possible because QCD is asymptotically free~\cite{Politzer:2005kc, Wilczek:2005az, Gross:2005kv}: asymptotic freedom guarantees that the scalar part of the quark self energy vanishes when Higgs boson couplings are removed~\cite[Secs.\,III, IV]{Maris:1997tm}.

Pursuing the chiral limit further, note that in this case the chromodynamics Lagrangian in Eq.\,\eqref{QCDdefine}, treated as defining a classical theory before field quantization, is invariant under scale transformations 
$x_\mu \to x'_\mu = \sigma x_\mu$, with $\sigma$ being a constant dilation factor.
This invariance is associated with the Noether dilation current 
$D_\mu = T_{\mu \nu} x_\nu$, where $T_{\mu \nu}$ is the chromodynamics energy--momentum tensor (EMT), which must then be conserved.
Thus, scale invariance is the statement $\partial_\mu D_\mu = 0 = T_{\mu\mu}$, \textit{viz}., the EMT must be traceless in a scale invariant theory~\cite{Roberts:2016vyn}.
Considering, then, the forward limit of the in-nucleon expectation value of the chromodynamics EMT, one has
\begin{equation}
0 = m_N \langle N(p) | T_{\mu\mu}^{\rm classical} = 0  | N(p) \rangle
= - p^2 = m_N^2\,;
\end{equation}
namely, the nucleon (and all other hadrons) must be massless. 

This is one consequence of the following general statements.
In scale-invariant theories, dynamics cannot be supported; only kinematics persists. 
Further, since all length scales are equivalent, there is no stabilization mechanism to support bound states. 
Confinement cannot be realized because quarks and gluons can be separated by any distance, and all such configurations would remain indistinguishable in scale-invariant theories.

After quantization, classical chromodynamics becomes QCD and a mass scale is introduced through the regularization and renormalization procedure, which is required to eliminate ultraviolet divergences and is characterized by a renormalization scale, $\zeta$.
(Infrared divergences may also appear in perturbation theory and can be treated with mathematical rigor.)
Naturally, true observables cannot depend on the choice of $\zeta$; however, many other intermediate, useful quantities do.  
For instance, the quantization procedure entails that the QCD coupling comes to depend on the energy scale: $\alpha \to \alpha(\zeta)$. 
The derivative of the logarithm of the QCD running coupling with respect to $\zeta$ defines the QCD $\beta$-function.
Following some straightforward algebra, one finds that, even in the chiral limit, the divergence of the QCD dilation current, $\partial_{\mu} D_{\mu}$, and, consequently, the trace of the QCD EMT, \(T_{\mu \mu}\), become non-zero~\cite{Roberts:2016vyn}:
\begin{equation}
\label{TraceAnomaly}
 \partial_\mu {\mathpzc D}_\mu =
\frac{\delta {\mathpzc L}_{\rm QCD}}{\delta \sigma} = \alpha \beta(\alpha) \, \frac{{\mathpzc L}_{\rm QCD}}{\delta \alpha} = \beta(\alpha) \tfrac{1}{4} G_{\mu\nu}^a G_{\mu\nu}^a = T_{\mu\mu}=: \Theta_0\,.
\end{equation}
This is an expression of the QCD trace anomaly.  

In QCD, therefore, the forward limit of the in-nucleon EMT expectation value is
\begin{equation}
m_N \langle N(p) | T_{\mu\mu} = \Theta_0 | N(p) \rangle
= - p^2 = m_N^2 \neq  0\,.
\end{equation}
Namely, so long as the QCD $\beta$-function is non-zero, gluons can potentially/probably generate a mass for the nucleon (and other hadrons), even in the limit of massless quarks.  The question is: How large is the trace anomaly contribution?

This is the crux.  
The existence of the trace anomaly merely indicates that a hadron mass scale is very likely generated by quantization of 4D QCD.
It provides no information nor insights regarding the underlying dynamics of hadron mass generation or the size of the effect.
A deeper understanding is needed to explain, \textit{e.g}., why the trace anomaly manifests itself so differently in generating pion and nucleon masses: at physical values of the light-quark current masses $m_\pi/m_N \approx 1/7$ and, in the chiral limit, $m_\pi/m_N =0$ (see Section~\ref{mb_structure}). 
Additionally, theoretical efforts should address the issue of predicting the value of the hadron mass scale, even though that is beyond-SM physics.

An extensive effort is underway at JLab to observe some/any manifestation of the trace anomaly in the in-nucleon expectation value of the EMT through experiments on $J/\psi$ photo- and electroproduction in the near-threshold region~\cite{Burkert:2023wzr}. 
These studies form an important part of the motivation for the experimental program at the Electron--Ion Collider (EIC)~\cite{Achenbach:2023pba}. 
However, numerous studies~\cite{Du:2020bqj, Xu:2021mju, Sun:2021pyw, JointPhysicsAnalysisCenter:2023qgg, Tang:2024pky, Sakinah:2024cza} have shown that significant improvement of reaction models is necessary before, if ever, it may become possible to deliver an objective extraction of nucleon gravitational form factors from $J/\psi$ photo- and electroproduction data and therewith shed light on the trace anomaly and a potential partonic gluon contribution to the proton mass.  
Today, any claim to have extracted a gluon contribution to the proton mass is insupportable. 

In the SM, the renormalization group invariant (RGI) current masses of the parton 
quarks are generated through the Higgs mechanism~\cite{Higgs:2014aqa, Englert:2014zpa}. 
A mechanism of this type was confirmed by discovery of the Higgs boson at CERN~\cite{ATLAS:2012yve, CMS:2012qbp}. 
It is responsible for generating the Lagrangian masses of the most fundamental constituents of matter known so far, namely, quarks and leptons. 
However, the Higgs mechanism plays a negligible role in generating the masses of nucleons and their excited states, $N^\ast$. 
This becomes evident when comparing the measured masses of protons and neutrons with the sum of the masses of their valence quark constituents, as presented in Table~\ref{nucleon_mass}. 
Protons and neutrons are bound systems seeded by three light quarks, \textit{i.e}., $uud$ and $udd$, respectively.
The sum of the current masses of these quarks, which is generated by Higgs couplings into QCD, accounts for less than 2\% of the measured nucleon masses. The comparison in Table~\ref{nucleon_mass} employs quark current masses from Ref.\,\cite[PDG]{PhysRevD.110.030001} at a scale $\zeta = 2.0\,$GeV. 
Use of RGI quark current masses leaves estimates for their contribution to the measured masses of the proton and neutron almost unchanged. 
This accounting indicates that the dominant part of the nucleon mass is created by mechanisms other than those associated with the Higgs boson.

\begin{table*}[t]
\begin{center}
\vspace{2mm}
\begin{tabular}{lcc} \toprule
                    & Proton          & Neutron \\ \midrule
\begin{tabular}{l} Measured masses (MeV) \end{tabular}     & 938.2720813     & 939.5654133 \\
 & $\pm$ 0.0000058 & $\pm$ 0.0000058 \\  \midrule           
  \begin{tabular}{l}
    Sum of the current \\
    quark masses (MeV)
  \end{tabular} &
  \begin{tabular}{l}
   8.09$^{+1.45}_{-0.65}$ 
  \end{tabular} &
  \begin{tabular}{l}
   11.50$^{+1.45}_{-0.60}$ 
  \end{tabular}
  \\ \midrule
\begin{tabular}{l} Contribution of the current \\
quark masses to the measured \\
nucleon mass (\%)
\end{tabular}  &  
\begin{tabular}{l} $<$ 1.1 
\end{tabular}  &
\begin{tabular}{l} $<$ 1.4 
\end{tabular} \\ \bottomrule
\end{tabular}
\end{center}
\caption{Comparison between the measured masses of the proton and neutron, $m_{N=p,n}$, and the sum of the current-quark masses of their three $u$ and $d$ valence quark constituents~\cite{PhysRevD.110.030001}. 
Current quark masses listed at a scale of 2\,GeV, but the comparison remains practically unchanged if renormalization group invariant current masses are used.
\label{nucleon_mass}}
\end{table*}

The past few decades have seen significant progress with experimental studies of the structure of pseudoscalar mesons and the ground and excited states of the nucleon~\cite{Horn:2016rip, Mokeev:2022xfo, Mokeev:2023zhq, Burkert:2022ioj, Proceedings:2020fyd, Carman:2016hlp}. 
Analyses of these results, conducted through synergistic efforts involving experiment, phenomenology, and QCD-based hadron structure theory, have conclusively demonstrated that the dominant part of a baryon's mass is generated by strong interactions at distance scales beyond that which marks the pQCD-to-sQCD transition boundary~\cite{Horn:2016rip, Roberts:2021nhw, Ding:2022ows, Carman:2023zke, Raya:2024ejx}.
Recent progress in theory has enabled this point to be located; see below.  


Working with CSMs, a unified theoretical framework has been established for elucidating the role of EHM in both the meson and baryon sectors, including ground and excited states of nucleons.  
In fact, currently, CSMs provide the only QCD-connected approach capable of predicting the $Q^2$ evolution of nucleon resonance electroexcitation amplitudes within the same framework used to explain the properties of many other hadrons, including Nature's most fundamental Nambu-Goldstone bosons.
Development of other nonperturbative theoretical tools for QCD to a point wherefrom they can supply a similar array of results is eagerly awaited.  

\begin{figure}[t]
\includegraphics[width=0.47\textwidth]{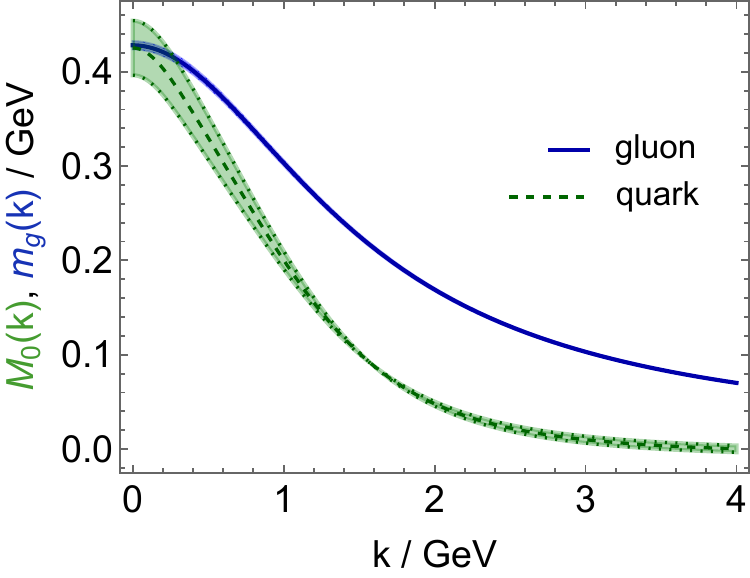}
\hspace*{1em}\includegraphics[width=0.48\textwidth]{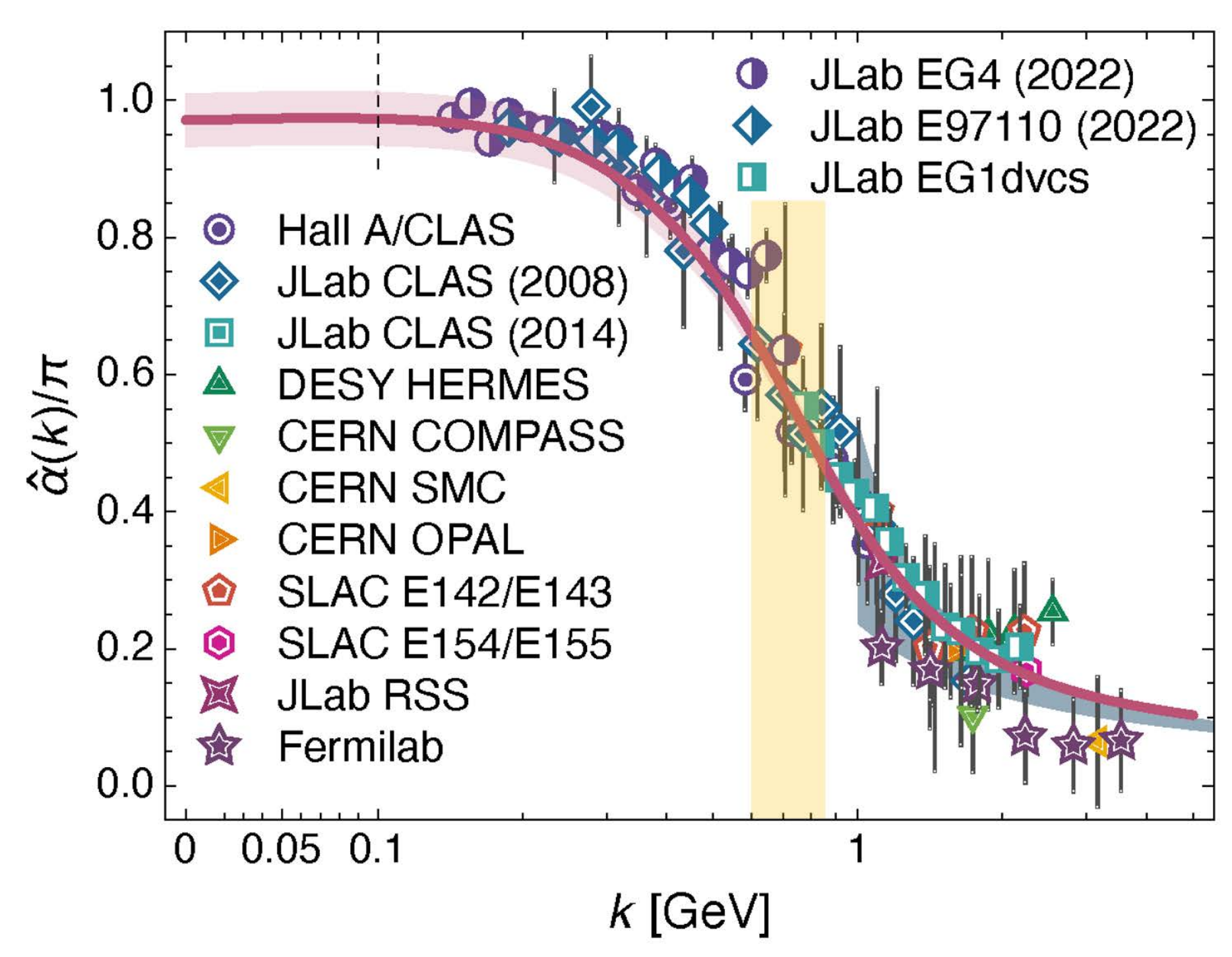}
\caption{\label{Quark_Gluon_massfunct}
Left: CSM predictions for the momentum dependence of the dressed gluon (solid blue curve) and quark (dot-dashed green) mass functions in the chiral limit~\cite{Roberts:2021xnz, Roberts:2020hiw, Roberts:2021nhw}.  
For the quark, the associated like-colored bands express existing uncertainties in the CSM predictions.  Relative uncertainties for the gluon are similar.
Crucially, both functions are essentially nonperturbative, \textit{viz}., neither can appear at any finite order in perturbation theory.
(\textit{N.B}.\ Since the Poincar\'e-invariant kinetic energy operator for a vector boson has mass-dimension two and that for a spin-half fermion has mass-dimension unity, then for $m_p^2/k^2 \to 0$, $M_0(k) \propto 1/k^2$ and $m_g^2(k) \propto 1/k^2$, up to $\ln k^2$ corrections). 
Right: CSM prediction~\cite{Cui:2019dwv} for the process-independent QCD running coupling, $\hat\alpha(k)$ (soft-purple curve and associated uncertainty band, which includes uncertainties associated with the gluon mass function) compared with empirical results~\cite{Deur:2022msf} for the process-dependent effective charge defined via the Bjorken sum rule.  
The vertical yellow band marks the window of sQCD $\leftrightarrow$ pQCD transition in the running coupling --- see Eq.\,\eqref{spWindow}.
Notably, in contrast to $M_0(k)$, $m_g(k)$, the running coupling is non-zero in pQCD: a one-loop approximation is reliable on $k\gtrsim 2\,m_0$.
(A complete discussion of effective charges is available elsewhere~\cite{Deur:2023dzc}.  All sources of the data in the right panel are listed in Refs.\,\cite{Deur:2022msf, Deur:2023dzc}.)}

\end{figure}

CSM studies predict that, owing solely to the gluon self-interactions evident in Eq.\,\eqref{QCDdefine}, which become strong at infrared scales, massless gluon partons undergo a remarkable metamorphosis, becoming quasiparticles characterized by a running mass function that is large at infrared momenta. 
This mass function, illustrated in Fig.\,\ref{Quark_Gluon_massfunct}\,-\,left, is entirely nonperturbative and the purest expression of QCD's trace anomaly and realization of EHM: massless gluons become massive as a direct consequence of interactions amongst themselves.  
Couched in RGI terms, the gluon mass scale is~\cite{Cui:2019dwv}:
\begin{equation}
\label{RGIGluonMass}
m_0 = 0.43(1)\,{\rm GeV}.
\end{equation}
No ``human interference'' is required here, \textit{i.e}., no Higgs boson phenomenology or anything like it is needed in order to engineer the appearance of a gluon mass scale.
The phenomenon is a consequence of a Schwinger mechanism in QCD~\cite{Schwinger:1962tn, Schwinger:1962tp, Cornwall:1981zr}, \textit{i.e}., a remarkable dynamical effect at work in QCD's gauge sector, which manifests itself in the generation of a mass scale in the gluon vacuum polarization.
A contemporary perspective on this physics is provided elsewhere~\cite{Binosi:2022djx, Ferreira:2023fva}.
Importantly, as explained therein, both CSM and lQCD analyses of QCD's gauge sector agree on these outcomes, which may therefore be described as QCD facts. 

There is no simple mapping between the partonic form of the trace anomaly in Eq.\,\eqref{TraceAnomaly} and the emergence of a gluon mass scale.  
Since the outcome is essentially nonperturbative, there cannot be. 
Nevertheless, as a Poincar\'e-invariant matrix element, the non-zero value of $\langle N(p) | \Theta_0 | N(p) \rangle$ is independent of the degrees of freedom used in its evaluation. 
Such a calculation has thus far proved impossible using a partonic basis.
On the other hand, using CSMs, with their exploitation of gap equations and the quasiparticles they produce, tractable calculations have become available; see, \textit{e.g}., Refs.\,\cite{Eichmann:2016yit, Qin:2019hgk, Yao:2024uej}.
They provide a direct means of connecting the gluon mass scale with $m_N$ and many other features of strong interaction observables.  

The emergence of a gluon mass scale has far-reaching consequences.  
One of the most fundamental is expressed in QCD's running coupling, a context for which is provided in Ref.\,\cite{Deur:2023dzc}.  
As explained in Ref.\,\cite{Binosi:2016nme}, using the pinch technique~\cite{Pilaftsis:1996fh, Binosi:2009qm, Cornwall:2010upa} and background field method~\cite{Abbott:1981ke}, it becomes possible in QCD to define and calculate a unique, process-independent (PI) and RGI analogue of the Gell-Mann--Low effective charge, first introduced for quantum electrodynamics (QED)~\cite{Gell-Mann:1954yli}.
The most detailed analysis of this charge, often denoted $\hat\alpha(k^2)$, is provided in Ref.\,\cite{Cui:2019dwv}.
Therein, modern results from continuum analyses of QCD's gauge sector and lQCD configurations generated with three domain-wall fermions at the physical pion mass~\cite{RBC:2014ntl, Boyle:2015exm, Boyle:2017jwu} were used to obtain a parameter-free prediction of $\hat\alpha(k^2)$. 
The resulting charge is shown in Fig.~\ref{Quark_Gluon_massfunct}\,-\,right. 
It was calculated using a momentum subtraction renormalization scheme, in which the RGI scale $\Lambda_{\rm QCD}= 0.52\,$GeV when $n_f=4$.  
Mathematically, one such RGI scale is necessary and sufficient to complete the analysis; so, one can equally trade $\Lambda_{\rm QCD}$ for the gluon mass: 
\begin{equation}
    \Lambda_{\rm QCD} = 1.21\,m_0\,.
\end{equation}

Regarding Fig.~\ref{Quark_Gluon_massfunct}\,-\,right, one observes a remarkable feature.  
Namely, whereas the perturbative running coupling in QCD exhibits a Landau pole, \textit{i.e}., diverges at $\Lambda_{\rm QCD}$, irrespective of the loop-order in the calculation --- see the discussion in Ref.\,\cite{Deur:2023dzc}, the PI charge is a smooth function on $k^2 \geq 0$: the emergence of the gluon mass scale eliminates the Landau pole.  
This feature emphasizes the role of EHM as expressed in Eq.~\eqref{RGIGluonMass}: the existence of $m_0 \approx m_N/2$ guarantees that long-wavelength gluons are screened, so play no dynamical role.
(This also resolves the Gribov ambiguity~\cite{Gao:2017uox}.)

Working with the PI charge, one can compute the associated $\beta$-function:
\begin{equation}
\beta_{\hat\alpha}(t) = \frac{d}{dt} \ln \hat\alpha({\rm e}^{2 t})\,,
\quad t=(1/2)\ln k^2/\Lambda_{\rm QCD}^2.
\label{QCDbeta}
\end{equation}
The maximum value of $|\beta_{\hat\alpha}(t)|$ occurs at
    $k_{\beta_{\rm max}} \approx 2.0 \, m_0$.
This is the point at which the PI-charge $\beta$-function is changing most rapidly.  
The derivative of the PI charge itself takes its largest magnitude on the neighborhood $k_{\hat\alpha'_{\rm max}} \simeq 1.4 \, m_0$.
Halfway between these two locations lies the PI-charge screening mass; namely, the point at which the perturbative coupling would diverge, realizing a Landau pole, but the PI charge reaches half its maximum value: $\hat\alpha((1.7\Lambda_{\rm QCD})^2)/\pi \approx 0.5$.  
The pion and kaon structure function studies in Ref.\,\cite{Cui:2020tdf} identified this screening mass with the hadron scale:
\begin{equation}
\zeta_{\cal H} = 1.7\,m_0\,,
\end{equation}
\textit{i.e}., the renormalization scale whereat valence quasiparticle degrees-of-freedom should be used to formulate and solve hadron bound state problems.  This notion is proving efficacious; see, \textit{e.g}., Refs.\,\cite{Ding:2022ows, Carman:2023zke, Raya:2024ejx, Yao:2024ixu, Xu:2024nzp}.

The above discussion enables one to identify a domain of sQCD $\leftrightarrow$ pQCD transition in QCD's running coupling: 
\begin{equation}
{\rm sQCD}_{\hat\alpha} \quad \leftarrow \quad 1.4 \lesssim k/m_0 \lesssim 2.0 \quad \rightarrow \quad {\rm pQCD}_{\hat\alpha}\,,
\label{spWindow}
\end{equation}
which is marked in Fig.~\ref{Quark_Gluon_massfunct}\,-\,right; see the vertical yellow band.
At the borders of this domain, one has
$\hat\alpha/\pi \approx 0.6, 0.4$, respectively.

Figure~\ref{Quark_Gluon_massfunct}\,-\,right also displays experimental measurements of the process-dependent charge, $\alpha_{g_1}(k^2)$~\cite{Deur:2022msf}, determined from the Bjorken sum rule~\cite{Bjorken:1966jh, Bjorken:1969mm}.
Evidently, there is good agreement between the two charges. 
This is partly explained by the fact that the Bjorken sum rule is an isospin non-singlet relation, which suppresses many dynamical contributions that might distinguish between the two charges. 
The charges are not identical; but, equally, on any domain for which perturbation theory is valid, they are nevertheless much alike:
\begin{equation}
\frac{\alpha_{g_1}(k^2)}{\hat\alpha(k^2)} \stackrel{k^2 \gg m_0^2}{=} 1+\frac{1}{20} \alpha_{\overline{\rm MS}}(k^2)\,,
\end{equation}
where $\alpha_{\overline{\rm MS}}$ is the textbook pQCD running coupling.  
At the charm quark current-mass, the ratio is $1.007$, \textit{viz}., indistinguishable from unity insofar as currently achievable precision is concerned.
Considering the other extreme, the Bjorken charge saturates to $\alpha_{g_1}(k^2=0)=\pi$; hence,
\begin{equation}
\frac{\alpha_{g_1}(k^2)}{\hat\alpha(k^2)} \stackrel{k^2 \ll m_0^2}{=} 1.03(4)\,.
\end{equation}
This discussion establishes that, for sound reasons in mathematics and physics, the process-dependent charge determined from the Bjorken sum rule is practically indistinguishable from the PI charge that emerges from QCD's gauge sector dynamics~\cite{Binosi:2016nme, Cui:2019dwv}.

Since $\hat\alpha(k^2)$ is process independent, it can serve numerous purposes and unify many observables. 
Consequently, $\hat\alpha$ is a good candidate for the long-sought running coupling that describes QCD interactions at \textit{all} accessible momentum scales~\cite{Dokshitzer:1998nz}.
Critically, the properties of $\hat\alpha(k^2)$ support the conclusion that QCD is actually a theory, \textit{viz}., a mathematically well-defined $D=4$ quantum gauge field theory. 
As such, QCD emerges as a viable tool for extending the SM by giving substructure to particles that today seem elementary. 

QCD's quark partons are also transmogrified, and their emergent properties can be revealed by solving the quark gap equation; see Ref.\,\cite{Binosi:2016wcx} for details.
The general solution of the gap equation takes the following form: 
\begin{equation}
    S(k) = Z(k^2)/[i\gamma\cdot k + M(k^2)]\,,
    \label{PropQuark}
\end{equation}
where $Z(k^2)$ is the quark wavefunction renormalization function and $M(k^2)$ is the RGI quark mass function.  
The gauge sector elements discussed above are key pieces in the gap equation's kernel~\cite{Binosi:2016wcx}.  
Solving the equation obtained therewith, working in the chiral limit, one obtains a dressed-quark mass function of the type drawn in Fig.\,\ref{Quark_Gluon_massfunct}\,-\,left.  
Lattice-regularized QCD delivers similar outcomes~\cite{Bowman:2005vx, Virgili:2022wfx}.
Hence, in the matter sector, too, sQCD dynamics produces mass from nothing: massless quark partons become massive owing to interactions with their own gluon quasiparticle field.  
Notably, $3\,M(0) \approx m_N$.
In this outcome, one sees a natural explanation for the scale of the nucleon mass and its connection with emergent features of QCD.

Here, it should be stressed that $M_0(k)$, $m_g(k)$ in Fig.~\ref{Quark_Gluon_massfunct}\,-\,left are essentially nonperturbative.  
So, whereas a pQCD approximation to the running coupling can be useful on $k\gtrsim 2 m_0$, that is not the case for the gluon and quark mass functions: they are both identically zero at all orders in perturbation theory. 
Regarding $M_0(k)$, a domain of sQCD dominance can be identified by comparing this function with its $u$-quark kin, $M_u(k)$, which is non-zero in pQCD.
Reviewing Ref.\,\cite[Fig.\,2.5]{Roberts:2021nhw}, $M_0(k)$ becomes clearly distinguishable from $M_u(k)$ on
\begin{equation}
    {\rm sQCD}_{M_0} \quad  \leftarrow \quad 5 \lesssim k/m_0 \,.
    \label{QCDM0}
\end{equation}
Evidently, the domain of sQCD dominance for the quark mass function extends far into the pQCD domain of the running coupling.  

Together, the emergence of a RGI gluon mass scale, $m_0$, 
the existence and infrared saturation of a process-independent QCD running coupling, $\hat\alpha(k^2)$, 
and the chiral-limit generation of dressed quarks with a RGI running mass, $M_0(k^2)$, constitute the three pillars of the EHM paradigm.  
Working with these elements, directly when possible and using well-constrained \textit{Ans\"atze} otherwise, one is readily able to deliver structure predictions for ground and excited state nucleons in a mass range that reaches, at least, to $2\,$GeV. 

Regarding $N^\ast$ electroexcitation, in the leading order of a systematic CSM approximation scheme, a virtual photon interacts, in turn, with each of the three dressed quarks that characterize the system. 
The amplitudes for such interactions are sensitive to the dressed-quark propagators; hence, a quark mass function of the type drawn in Fig.~\ref{Quark_Gluon_massfunct}\,-\,left.
Consequently, CSM analyses of the $Q^2$ evolution of $N^\ast$ electroexcitation amplitudes deliver empirical insights into the momentum dependence of the dressed quark mass. 
The range of coverage increases with $Q^2$~\cite{Ding:2022ows, Carman:2023zke, Proceedings:2020fyd}, scanning the infrared at low $Q^2$ and reaching beyond the sQCD$_{M_0}$ boundary as $Q^2$ is increased into the domain on which pQCD effects are dominant.

Empirical information on $N^\ast$ electroexcitation amplitudes, obtained from exclusive meson electroproduction data, is continually expanding. 
In the near term, these results are expected to become available for most $N^\ast$ states with masses up to 2~GeV and $Q^2 < 5$~GeV$^2$, based on data from the 6-GeV era with CLAS. 
Furthermore, experiments with CLAS12, currently underway in Hall~B at JLab, aim to extend these measurements up to $Q^2 \approx 10\,$GeV$^2$~\cite{Mokeev:2024beb}.  

Available and forthcoming results on $N^\ast$ electroexcitation amplitudes will enable the exploration of $Q^2$-scales that cover evolution of the dressed-quark mass function from its maximum value, $M(0)$, out to $M(0)/2$, \textit{i.e}., $0< k \lesssim k_{\beta_{\rm max}}$ or lengths $\gtrsim 0.2$\,fm.
Moreover, studies of the electroexcitation amplitudes for nucleon resonances of different structure provide a unique opportunity to establish either the universality or environmental sensitivity of the dressed quark mass function. These analyses also facilitate exploration of the connection between EHM and dynamical chiral symmetry breaking (DCSB), especially through the electroexcitation amplitudes of chiral partner resonances.  
Importantly, too, they will enable the examination of diquark correlations with different $J^P_{qq}$ spin-parities, as revealed in experimental results on the electroexcitation amplitudes of resonances with various spin-parity $J^{P}$. 
This will provide a window onto the partial wave decomposition of the scattering amplitudes for the dressed quarks. 

\section{Nucleon Resonance Electroexcitation Data from the JLab 6-GeV Era}

In recent decades, the information on the $Q^2$ evolution of $N^\ast$ electroexcitation amplitudes has been significantly expanded through measurements of exclusive meson electroproduction in the resonance region~\cite{Burkert:2022ioj,Burkert:2017djo,Mokeev:2022xfo,Carman:2023zke}. 
Analyses of exclusive meson electroproduction observables using various reaction models have enabled the determination of $N^\ast$ electroexcitation amplitudes for most nucleon resonances in the mass range up to 1.75\,GeV across a wide range of $Q^2$ from the photon point to 5~GeV$^2$. 
These experimental results provide critical input needed to gain insights into the strong interaction dynamics underlying EHM. 
This section highlights the advances in the study of the $N^\ast$ electroexcitation amplitudes achieved using data from the 6-GeV era at JLab. 

\subsection{Extraction of \texorpdfstring{$\gamma_vpN^\ast$}{Electrocouplings} Electrocouplings from Exclusive Electroproduction Data}

Nucleon resonance electroexcitation can be fully described by three $\gamma_v pN^\ast$ electrocouplings:  
\begin{equation}
\label{res_elcoupl}
A_{1/2}(Q^2), \qquad A_{3/2}(Q^2), \qquad S_{1/2}(Q^2).
\end{equation}  
The first two electrocouplings, $A_{1/2}(Q^2)$ and $A_{3/2}(Q^2)$, describe resonance electroexcitation by transversely polarized photons, while the third, $S_{1/2}(Q^2)$, corresponds to $N^\ast$ electroexcitation by longitudinally polarized photons, which is relevant only for the electroproduction process. 
All three electrocouplings depend on $Q^2$.  

These electrocouplings are proportional to the helicity amplitudes that describe the transition $\gamma_v + p \to N^\ast$ in the center-of-mass (CM) frame of the virtual photon and proton for different helicities of the initial particles. 
The subscripts in Eq.\,\eqref{res_elcoupl} correspond to the sum of the projections of the virtual photon and target proton (nucleon) spins onto an axis that is aligned along the three-momentum of the virtual photon. 
These electrocouplings are uniquely determined through their relationship with the resonance electromagnetic decay widths, $\Gamma_\gamma^T$ and $\Gamma_\gamma^L$, to the $\gamma_v p$ states with transversely and longitudinally polarized photons, respectively:

\begin{subequations}
\begin{align}
\Gamma_\gamma^T(W=M_r,Q^2) & =\frac{q^2_{\gamma,r}(Q^2)}{\pi}\frac{2M_N}{(2J_r+1)M_r} 
\left(|A_{1/2}(Q^2)|^2+|A_{3/2}(Q^2)|^2\right) , \\
\Gamma_\gamma^L(W=M_r,Q^2) & =\frac{q^2_{\gamma,r}(Q^2)}{\pi}\frac{2M_N}{(2J_r+1)M_r}|S_{1/2}(Q^2)|^2,\label{Eq:EMWidths}
\end{align}
\end{subequations}
with $q_{\gamma,r}=\left.q_{\gamma} \right|_{W=M_r}$ being the absolute value of the $\gamma_v$ three-momentum at the resonance point, $M_r$ and $J_r$ the resonance mass and spin, respectively, and $M_N$ the nucleon mass. $W$ is the sum of the energies of the $\gamma_v$ and target proton in their CM frame.

Alternatively, resonance electroexcitation can be described by three transition form factors, $G^\ast_{1,2,3}(Q^2)$ or $G^\ast_{M,E,C} (Q^2)$~\cite{Aznauryan:2011qj}, which represent Poincar\'e-invariant functions in the most general expressions for the $N \to N^\ast$ electromagnetic transition currents. 
For spin-1/2 resonances the number of form factors in the $N \to N^\ast$ transition currents reduces to two. 
Consequently, the $F^\ast_{1,2}(Q^2)$ Dirac and Pauli transition form factors can be used. 
The description of resonance electroexcitation in terms of the electrocouplings and the electromagnetic transition form factors is fully equivalent, as they are unambiguously related; see, {\it e.g.}, Refs.\,\cite{Aznauryan:2011qj, Obukhovsky:2019xrs}.

\begin{figure}[t]
\includegraphics[width=0.85\textwidth]{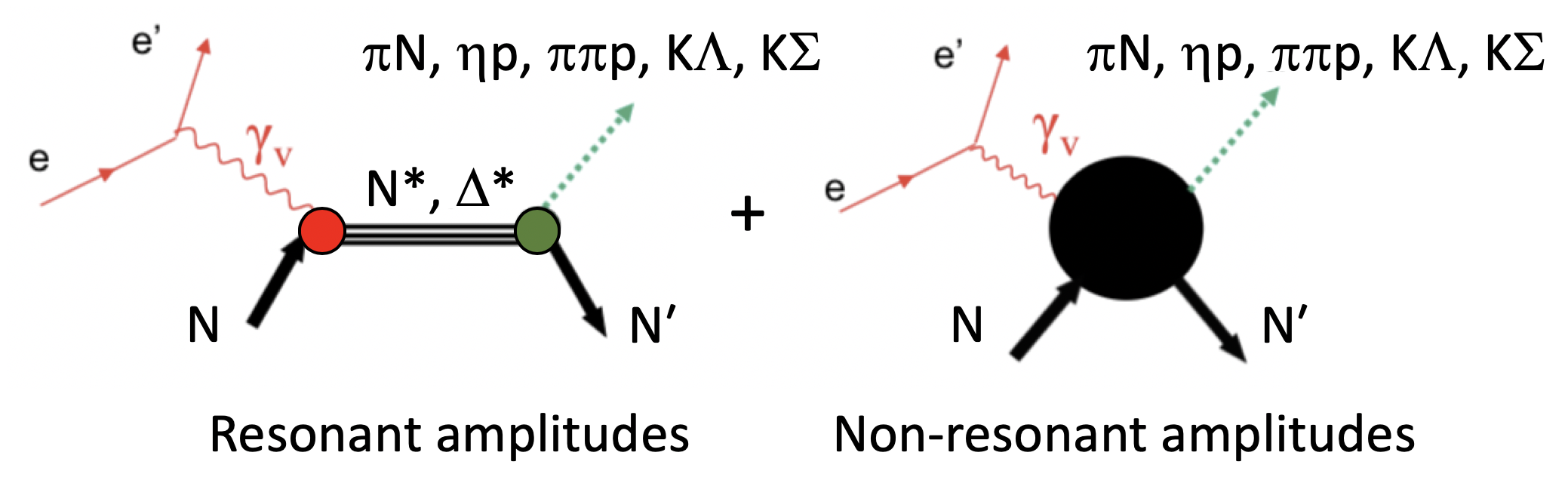}
\caption{Simplified diagrammatic representation of the resonant and non-resonant amplitudes contributing to various exclusive meson electroproduction channels in the resonance region.}
\label{exclusive_reactions1}
\end{figure}

The $\gamma_v pN^\ast$ electrocouplings have been extracted from observables of $\pi N$, $\eta p$, and $\pi^+\pi^- p$ electroproduction off protons. 
Independent and combined studies of the most relevant exclusive meson electroproduction channels in the resonance region are crucial for the reliable evaluation of these electrocouplings. 
The full amplitudes that describe any given electroproduction channel represent a superposition of $N^\ast$ electroexcitation in the $s$-channel for the virtual photon $+$ proton interaction, along with a complex set of amplitudes representing contributions other than $s$-channel resonance electroexcitation, known as the non-resonant contribution or background.
These different contributions are illustrated in Fig.\,\ref{exclusive_reactions1}.

The non-resonant amplitudes in various exclusive channels are entirely different. 
However, the $\gamma_v pN^\ast$ electrocouplings should be the same, as the $N^\ast$ electroexcitation amplitudes and their hadronic decay amplitudes to different final hadron states are independent. 
These amplitudes are separated in spacetime by the $N^\ast$ or $\Delta^\ast$ propagators; see Fig.\,\ref{exclusive_reactions1}\,-\,left. 
Therefore, consistent results for the $\gamma_v pN^\ast$ electrocouplings, obtained from independent analyses of different exclusive electroproduction channels, provide strong evidence for the capabilities of the reaction models to reliably extract the electrocouplings. 
The variation in the extracted values of these electrocouplings from such analyses allows for the evaluation of the systematic uncertainties associated with the use of reaction models in their extraction.

\subsection{Current Status of \texorpdfstring{$\gamma_v pN^\ast$}{} Electrocoupling Results}

Studies of nucleon resonance electroexcitation in various meson electroproduction channels became feasible following the availability of data from the 6-GeV era at JLab, collected in Hall~B using the CLAS detector and in Halls A and C. 
The majority of the world’s information on single and multi-meson electroproduction data off protons in the resonance region has been obtained using the CLAS detector for $Q^2$ up to 5~GeV$^2$~\cite{Burkert:2022ioj,Mokeev:2022xfo}.

\begin{table}[t]
\begin{center}
\vspace{2mm}
\begin{tabular}{ccccc} \toprule
Hadron Final & $W$ Coverage  & $Q^2$ Coverage & Measured &  References \\
State        & (GeV)            & (GeV$^2$)        & Observables&  \\ \midrule
$\pi^+ n$    &  1.1 -- 1.38      & 0.16 -- 0.36  & $\frac{d\sigma}{d\Omega}$& \cite{Smith:2007zzc}  \\
              & 1.1 -- 1.55      & 0.3 -- 0.6    & $\frac{d\sigma}{d\Omega}$ & \cite{CLAS:2006sjw}  \\
              & 1.1 -- 1.70      & 1.7 -- 4.5    & $\frac{d\sigma}{d\Omega}$, $A_b$ & \cite{CLAS:2007jpl}  \\ 
              & 1.1 -- 1.66      & 0.4 -- 0.65   &  $A_{bt}$ & \cite{CLAS:2004ncx}  \\  
              & 1.6 -- 2.00      & 1.8 -- 4.5    & $\frac{d\sigma}{d\Omega}$ & \cite{CLAS:2014fml}   \\ \midrule
$\pi^0 p$    &  1.1 -- 1.38      & 0.16 -- 0.36  & $\frac{d\sigma}{d\Omega}$ & \cite{Smith:2007zzc} \\
              & 1.1 -- 1.68      & 0.4 -- 1.8    & $\frac{d\sigma}{d\Omega}$, $A_b$, $A_t$, $A_{bt}$& \cite{CLAS:2001cbm, CLAS:2003vka, CLAS:2008wls} \\
              & 1.1 -- 1.39      & 3.0 -- 6.0    & $\frac{d\sigma}{d\Omega}$& \cite{CLAS:2001cbm}  \\
              & 1.1 -- 1.38      & 0.16 -- 0.36    & $\frac{d\sigma}{d\Omega}$& \cite{Smith:2007zzc}  \\           
              & 1.1 -- 1.80      & 0.4 -- 1.0    & $\frac{d\sigma}{d\Omega}$, $A_b$& \cite{CLAS:2019cpp, CLAS:2021cvy}    \\  \midrule   
$\eta p$      & 1.5  -- 2.30      & 0.2 -- 3.1    & $\frac{d\sigma}{d\Omega}$ & \cite{CLAS:2007bvs}   \\  
             & 1.50 -- 1.80      & 5.7 -- 7.0    & $\frac{d\sigma}{d\Omega}$ & \cite{Villano:2009sn}   \\
             & 1.49 -- 1.62      & 2.4 -- 3.6    & $\frac{d\sigma}{d\Omega}$ & \cite{Frolov:1998pw}   \\  \midrule         
$K^+ \Lambda $      & 1.61 -- 2.60      & 1.40 -- 3.90    & $\frac{d\sigma}{d\Omega}$ & \cite{CLAS:2006ogr,Carman:2012qj}  \\ 
                    &                & 0.5 -- 3.90     & $A_b$ & \cite{CLAS:2008agj,Carman:2012qj} \\
                    &                & 0.70 -- 5.40    & $P^0$, $P'$ & \cite{CLAS:2014udv,CLAS:2022yzd}  \\ \midrule
$K^+ \Sigma^0 $     & 1.68 -- 2.60      & 1.40 -- 3.90    & $\frac{d\sigma}{d\Omega}$ & \cite{CLAS:2006ogr,Carman:2012qj}  \\ 
                    &                & 1.40 -- 3.90    & $A_b$ & \cite{,Carman:2012qj} \\
                    &                & 0.70 -- 5.40    & $P'$ &  \cite{CLAS:2022yzd}  \\ \midrule    
$\pi^+ \pi^- p$     &  1.3 -- 1.6       & 0.20 -- 0.60  & Nine 1-fold & \cite{CLAS:2008ihz} \\
              & 1.4 -- 2.10      & 0.4 -- 1.5    & differential cross & \cite{CLAS:2002xbv, CLAS:2018fon}   \\
              & 1.4 -- 2.00      & 2.0 -- 5.0    & sections & \cite{CLAS:2017fja, Trivedi:2018rgo}  \\  \bottomrule
\end{tabular}
\end{center}
\caption{Summary of the exclusive meson electroproduction data obtained in the resonance region in experiments during the 6-GeV era at JLab and used for extraction of the $\gamma_v pN^\ast$ electrocouplings, including differential cross sections $\frac{d\sigma}{d\Omega}$, beam $A_b$, target $A_t$, and beam--target $A_{bt}$ asymmetries, and recoil $P^0$ and transferred $P'$ polarizations of $\Lambda$ and $\Sigma^0$ hyperons.
\label{exclusive_chann_obser}}
\end{table}

The data are stored in the CLAS Physics Database~\cite{CLAS:DB,Chesnokov:2022gjb} and in the SAID database~\cite{saiddb}. 
For the first time, an extensive dataset ($\approx 150,000$ points) on differential cross sections and polarization asymmetries has been made available, with nearly complete coverage of the final-state hadron CM-emission angles. 
This coverage is critical for reliable extraction of the electrocouplings. 
A summary of the meson--baryon electroproduction data obtained from the 6-GeV era experiments at JLab and used to extract the $\gamma_v pN^\ast$ electrocouplings is presented in Table~\ref{exclusive_chann_obser}.

Several approaches have been developed for the extraction of the electrocouplings from these data that were determined from independent studies of the individual channels: 
$\pi^+ n$ and $\pi^0 p$~\cite{Tiator:2018pjq, Tiator:2011pw, Aznauryan:2002gd, CLAS:2009ces, CLAS:2014fml}, 
$\eta p$~\cite{Aznauryan:2003zg, Knochlein:1995qz, CLAS:2007bvs}, 
and $\pi^+\pi^- p$~\cite{Mokeev:2008iw, CLAS:2012wxw, Mokeev:2015lda, Mokeev:2023zhq}. 
The development of a global multi-channel analysis for $\pi N$, $\eta N$, $K\Lambda$, and $K\Sigma$ photo-, electro-, and hadroproduction to extract the $\gamma_v pN^\ast$ electrocouplings within an advanced coupled-channels approach by the J{\"u}lich--Bonn--Washington group represents an important breakthrough~\cite{Wang:2024byt, Mai:2021aui}. 
For the first time, this approach allows a complementary determination of the electrocouplings for most $N^\ast$ states up to 1.75~GeV for $Q^2 < 5$~GeV$^2$ from the available data of the abovementioned channels in a combined analysis that accounts for the final state interactions described within the models that are checked in comparison with the data of experiments with hadronic probes. 
Earlier results on the electrocouplings of the $\Delta(1232)3/2^+$ and $N(1440)1/2^+$ were obtained from a global multi-channel analysis of photo-, electro-, and hadroproduction data within the coupled-channels approach developed by the Argonne--Osaka group~\cite{Kamano:2018sfb}. The results on the $\gamma_v pN^\ast$ electrocouplings obtained from independent studies of the $\pi N$ and $\pi^+\pi^-p$ electroproduction channels are further supported by a coupled-channels analysis conducted by the J{\"u}lich--Bonn--Washington group~\cite{Wang:2024byt}. 

Most electrocoupling results to date have been derived from independent analyses of CLAS data on $\pi N$ and $\pi^+\pi^-p$ electroproduction. 
The unitary isobar model and dispersion-relation approaches developed by the CLAS Collaboration for $\pi N$ electroproduction~\cite{Aznauryan:2002gd, CLAS:2009ces, CLAS:2014fml} provide a good description of the observables for $W$ up to 1.7~GeV and $Q^2 < 5$~GeV$^2$. 
Similarly, the data-driven JLab-Moscow (JM) meson--baryon reaction model successfully describes the $\pi^+\pi^-p$ electroproduction data for $W < 2$~GeV and $Q^2 < 5$~GeV$^2$~\cite{CLAS:2012wxw, Mokeev:2015lda, Mokeev:2023zhq}. 

These reaction models for $\pi N$ and $\pi^+\pi^-p$ electroproduction enable the reliable isolation of the resonant contributions, essential for extracting the electrocouplings. 
The $\gamma_v pN^\ast$ electrocouplings were deduced from the isolated resonant amplitudes accounting for restrictions imposed by the unitarity condition on the resonant contributions~\cite{CLAS:2012wxw, Aznauryan:2011qj}. A summary of the results on the $\gamma_v pN^\ast$ electrocouplings obtained from the exclusive meson electroproduction data measured with CLAS is presented in Table~\ref{electrocoupl_sum}. 
The most recent results are reported in Refs.\,\cite{HillerBlin:2019jgp, Mokeev:2020hhu, Mokeev:2023zhq, Mokeev:2024beb}.

\begin{table}[t]
\begin{center}
\vspace{2mm}
\begin{tabular}{c|cc} 
\toprule
Channel        & Excited Nucleon & $Q^2$ Range (GeV$^2$) of \\
               & States          & Electrocouplings \\
\midrule
\multirow{2}{*}{$\pi^+ n$, $\pi^0p$}    & $\Delta(1232)3/2^+$,       & 0.16 -- 6.0  \\ [-2.1ex]\\ \cline{2-3} & \\[-1.8ex]
                                        & $N(1440)1/2^+$, $N(1520)3/2^-$, $N(1535)1/2^-$ & 0.3 -- 4.16 \\
\midrule
$\pi^+n$       & $N(1675)5/2^-$, $N(1680)5/2^+$, $N(1710)1/2^+$ & 1.6 -- 4.5 \\ 
\midrule
$\eta p$       & $N(1535)1/2^-$                                 & 0.2 -- 2.6 \\ 
\midrule
\multirow{3}{*}{$\pi^+ \pi^-p$} & $N(1440)1/2^+$, $N(1520)3/2^-$, $\Delta(1600)3/2^+$,       & 0.25 -- 5.0  \\ [-2.1ex]\\ \cline{2-3} & \\[-1.8ex]
               & $\Delta(1620)1/2^-$, $N(1675)5/2^-$, $N(1680)5/2^+$,  & \multirow{2}{*}{0.5 -- 1.5}  \\ 
               &  $\Delta(1700)3/2^-$, $N(1720)3/2^+$, $N'(1720)3/2^+$       &   \\ 
\bottomrule
\end{tabular}                        
\end{center}
\caption{The $\gamma_vpN^\ast$ electrocouplings determined from independent analyses of different meson electroproduction channels measured with CLAS.
\label{electrocoupl_sum}}
\end{table}

\begin{figure}[t]
\includegraphics[width=0.85\textwidth]{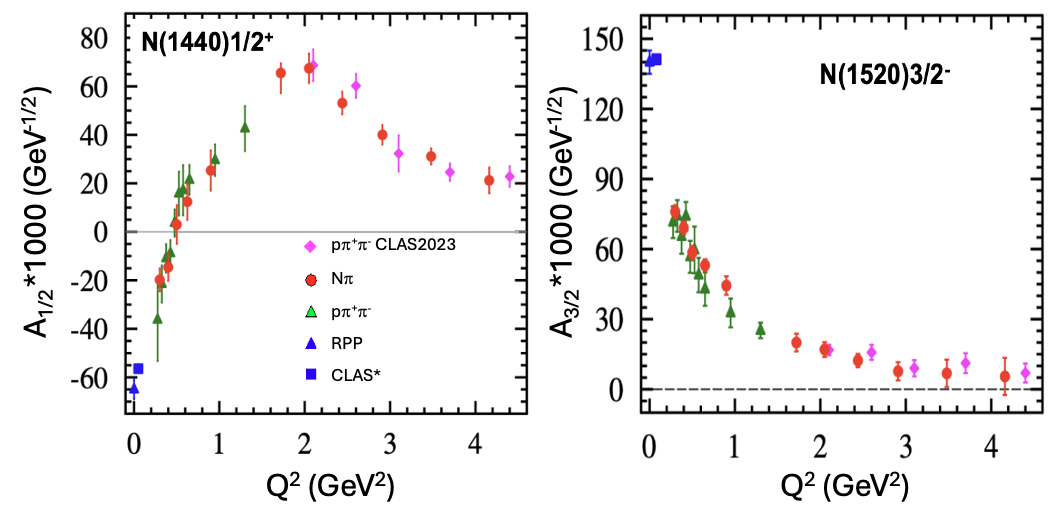}
\vspace{-1.5mm}
\caption{$N(1440)1/2^+$ and $N(1520)3/2^-$ electrocouplings extracted from the $\pi N$~\cite{CLAS:2009ces} and $\pi^+\pi^-p$~\cite{Mokeev:2015lda, CLAS:2012wxw, Mokeev:2023zhq} electroproduction channels. The photocouplings from the Particle Data Group (PDG)~\cite{PhysRevD.110.030001} and from Ref.\,\cite{CLAS:2009tyz} are shown by the blue squares and triangles, respectively. 
The key to the symbols and the color coding is shown in the left panel.
\label{p11d13_elcoupl}}
\end{figure}

Figure~\ref{p11d13_elcoupl} shows representative examples of the transverse $A_{1/2}(Q^2)$ electrocoupling for the $N(1440)1/2^+$ and the $A_{3/2}(Q^2)$ electrocoupling for the $N(1520)3/2^-$ versus $Q^2$, obtained from independent studies of the $\pi N$~\cite{CLAS:2009ces,CLAS:2014fml} and $\pi^+\pi^-p$~\cite{Mokeev:2015lda, CLAS:2012wxw, Mokeev:2023zhq} channels. 
The electrocouplings inferred from data on the two major $\pi N$ and $\pi^+\pi^-p$ electroproduction channels, with different non-resonant contributions, are consistent. 
This success exemplifies the capabilities of the reaction models developed by the CLAS Collaboration for the credible extraction of the $\gamma_vpN^\ast$ electrocouplings from independent studies of the $\pi N$ and $\pi^+\pi^-p$ electroproduction channels for nucleon resonances located within the mass range up to 1.6\,GeV.

\begin{figure}[t]
\includegraphics[width=0.85\textwidth]{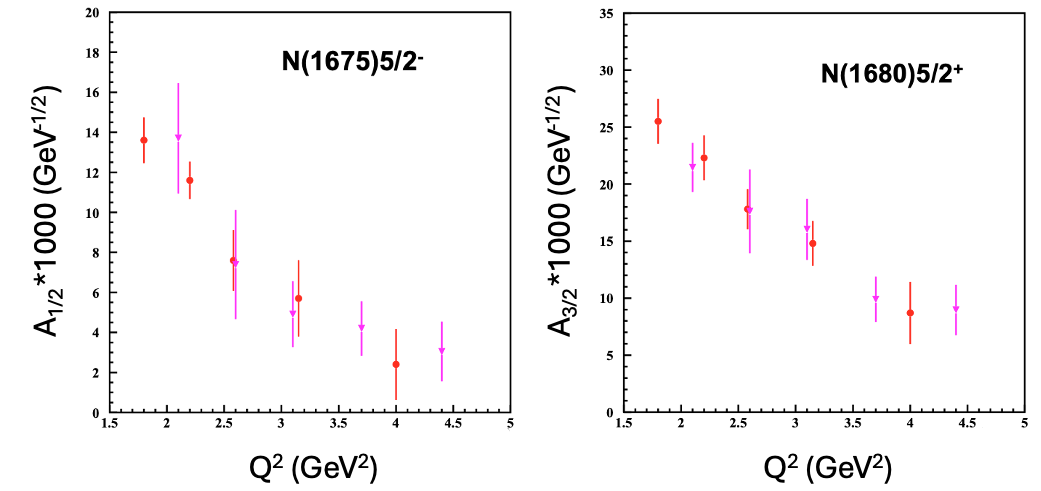}
\vspace{-1.5mm}
\caption{$N(1675)5/2^-$ and $N(1680)5/2^+$ electrocouplings extracted from the $\pi N$~\cite{CLAS:2014fml} and $\pi^+\pi^-p$~\cite{mokeev-nstar24} electroproduction channels. The key to the symbols and the color coding is the same as in Fig.~\ref{p11d13_elcoupl}.}
\label{d15f15_elcoupl}
\end{figure}

Similarly, the capabilities of the aforementioned reaction models have been verified for extracting the $\gamma_v pN^\ast$ electrocouplings from the $\pi N$ and $\pi^+\pi^-p$ electroproduction channels for resonances in the third resonance region. 
The comparison of the electrocouplings for the $N(1675)5/2^-$ and $N(1680)5/2^+$ is shown in Fig.\,\ref{d15f15_elcoupl}. 
The consistent results for both resonances across the range of $Q^2$ from 2 -- 5\,GeV$^2$ covered by the CLAS measurements conclusively demonstrate the reliability of the developed reaction models for extracting the electrocouplings in the third resonance region.

\section{Insight into EHM from Studies of \boldmath \texorpdfstring{$N^\ast$}{Nucleon Resonance} Structure}
\label{EHM_res_elcoupl}

This section explores new opportunities for mapping the momentum dependence of the dressed quark mass from results on the $Q^2$ evolution of the $\gamma_v pN^\ast$ electrocouplings.

\subsection{Dressed Quark Running Mass and its Connection to \texorpdfstring{$N^\ast$}{Nucleon Resonance} Structure}
We begin by discussing the running mass of a dressed quark. 
Owing to confinement, a dressed quark is always off-shell.
Nonetheless, it is worth recalling the mass definition of a particle on-shell: the Poincar\'e-invariant mass, $M$, of an on-shell particle is:
\begin{equation}
M^2 = E^2 - \vec{p}^2,
\label{mass_onshell}
\end{equation}
where $E$ and $\vec{p}$ represent the particle's energy and three-momentum, respectively, as usual. 

To describe propagation of an off-shell particle, which is of relevance for confined quarks, the definition of mass must be generalized. 
In quantum field theory (QFT), the particle mass, $M$, can be determined via the propagator. 
In Euclidean space, this is a two-point Schwinger function.  
For a spin-1/2 particle, such a propagator is proportional to:
\begin{eqnarray}
\frac{1}{p^2+M^2} \;\;\;\mbox{with}\; \;\;
p^2=(iE)^2+\vec{p}^2.
\label{mass_propag}
\end{eqnarray}
Note, in general, $p^2\neq -M^2$, so Eq.\,\eqref{mass_propag} is often interpreted as describing the propagation of a particle with non-zero virtuality, \textit{i.e}., an off-shell particle.
The poles in Eq.\,\eqref{mass_propag}, \textit{viz}., solutions of $p^2+M^2 = 0$, are associated with the on-shell (directly measurable) mass of the particle.  

Now recall Eq.\,\eqref{PropQuark}. This is the general form for the Euclidean propagator of any spin-1/2 object in QFT.  
Focusing on QCD, the RGI running quark mass depends on the quark's squared four-momentum: $k^2=(iE)^2+\vec{k}^2 = k_4^2+\vec{k}^2 > 0$ in Euclidean space.
The images in Fig.\,\ref{Quark_Gluon_massfunct}\,-\,left depict running masses as a function of $k=\sqrt{k^2}$.
The mass function in the dressed-quark propagator, $M(k)$ or $M(k^2)$, can be accessed via experimental results on the $Q^2$ evolution of the $\gamma_v pN^\ast$ electrocouplings.

\begin{figure*}[t]
\begin{center}
\includegraphics[width=1.0\textwidth]{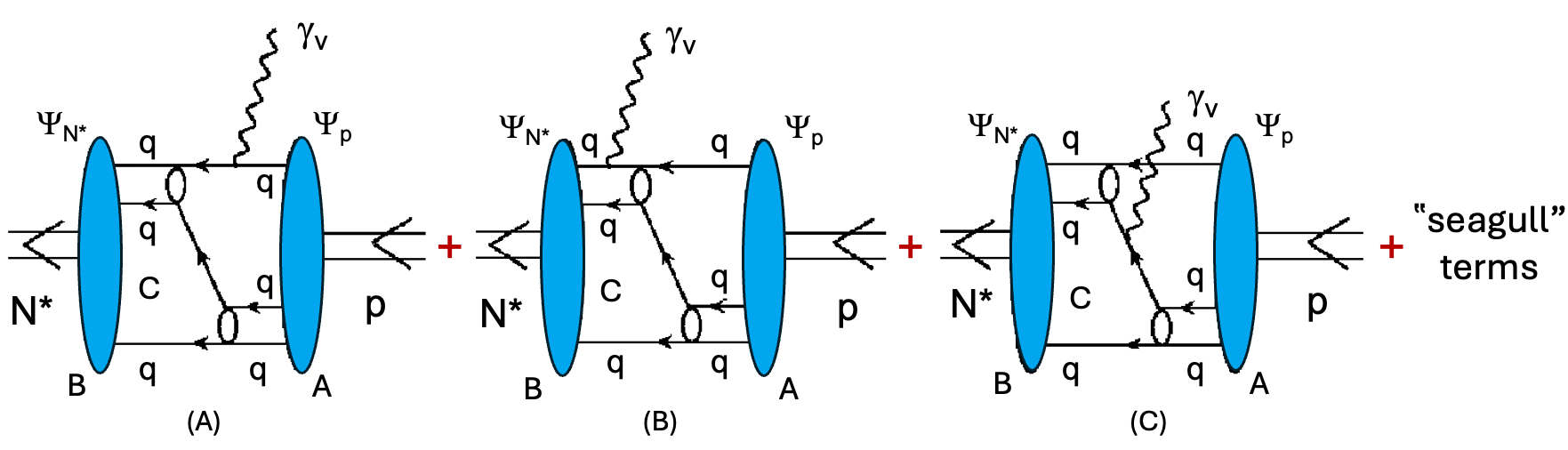}
\vspace{-0.1cm}
\caption{CSM description of resonance electroexcitation amplitudes~\cite{Cloet:2013jya, Segovia:2014aza, Segovia:2015hra, Burkert:2017djo, Barabanov:2020jvn}. 
All diagrams describe the transition $p \to$ dressed quark plus fully interacting diquark correlations $\to$ $N^\ast$. 
The Faddeev amplitudes for the transitions between dressed quark + diquark configurations and the ground or excited states of the nucleon correspond to the proton, $\psi_p$, or $N^\ast$, $\psi_{N^\ast}$, wavefunctions, respectively. 
The virtual photon interacts with all quark propagators as shown in part in panels (A) and (C). It also interacts with the diquark correlations as shown in panel (B) and to the transition diquark $\to$ $q+q$ amplitudes shown by the open ovals. 
The full set $N \to N^\ast$ transition amplitudes can be found in, \textit{e.g}., Ref.\,\cite[Fig.\,C1]{Segovia:2014aza}.
\label{diagdse}} 
\end{center}
\end{figure*}

Using CSMs, nucleon resonance electroexcitation is described by an electromagnetic current conserving set of diagrams~\cite{Cloet:2013jya, Segovia:2014aza, Segovia:2015hra, Burkert:2017djo, Barabanov:2020jvn}, partially depicted in Fig.~\ref{diagdse}. 
They represent the dominant contributions to the resonance electroexcitation amplitudes.
The full set of transition amplitudes can be found in Ref.\,\cite[Fig.\,C1]{Segovia:2014aza}.
The amplitudes describing the interactions of real or virtual photons with dressed quarks and diquark correlations are sensitive to the dressed quark propagators. 
The mass function of the dressed quark is a key component of the propagator, as defined by Eq.\,\eqref{PropQuark}. 
Consequently, studying the $Q^2$ evolution of the $\gamma_v pN^\ast$ electrocouplings provides a valuable opportunity to map out the momentum dependence of the dressed quark mass.

\subsection{Dressed Quark Mass Function Insights from \texorpdfstring{$\gamma_vpN^\ast$}{} Electrocouplings of 6-GeV Era at JLab}
\label{6gev_baryons}

Studies of the $\gamma_v pN^\ast$ electrocouplings, determined from exclusive meson electroproduction data measured with the CLAS detector in Hall B at JLab, have revealed that nucleon resonance structure for all investigated excited states of the nucleon represents a complex interplay between an inner core of three confined dressed colored quarks and an external meson--baryon cloud composed of colorless de-confined hadrons~\cite{Mokeev:2015lda, Aznauryan:2018okk, Burkert:2017djo, Mokeev:2022xfo, Burkert:2004sk, Suzuki:2010yn}. 
A general feature observed in $N^\ast$ electroexcitation is the gradual transition from the combined contributions of both the quark core and the meson--baryon cloud to quark-core dominance as $Q^2$ increases. 
The rate of this transition with $Q^2$ depends on the structure of the particular resonance.

This transition --- from interactions between quarks within the $N^\ast$ quark core to interactions between virtual mesons and baryons outside the quark core --- reflects the transition between the quark-gluon confinement and meson-baryon strong interaction regimes. 
These regimes underlie the generation of the two components in $N^\ast$ structure. 
High-virtuality photons penetrate the external meson--baryon cloud and predominantly interact with the quark core. 
For most $N^\ast$s, the quark core contributions become dominant for $Q^2$ above 1 -- 2~GeV$^2$. 
Thus, studies of the $Q^2$ evolution of the $\gamma_v pN^\ast$ electrocouplings are critical for understanding the interplay between meson--baryon and quark degrees of freedom in $N^\ast$ structure. 
Currently, the CSM approach accounts solely for contributions from the dressed quark degrees of freedom, as illustrated in Fig.\,\ref{diagdse}. 
A direct comparison between CSM predictions and experimental results for the $Q^2$ evolution of the $\gamma_v pN^\ast$ electrocouplings is meaningful only at length scales where quark core contributions dominate.

\begin{figure}[t]
\includegraphics[width=0.85\textwidth]{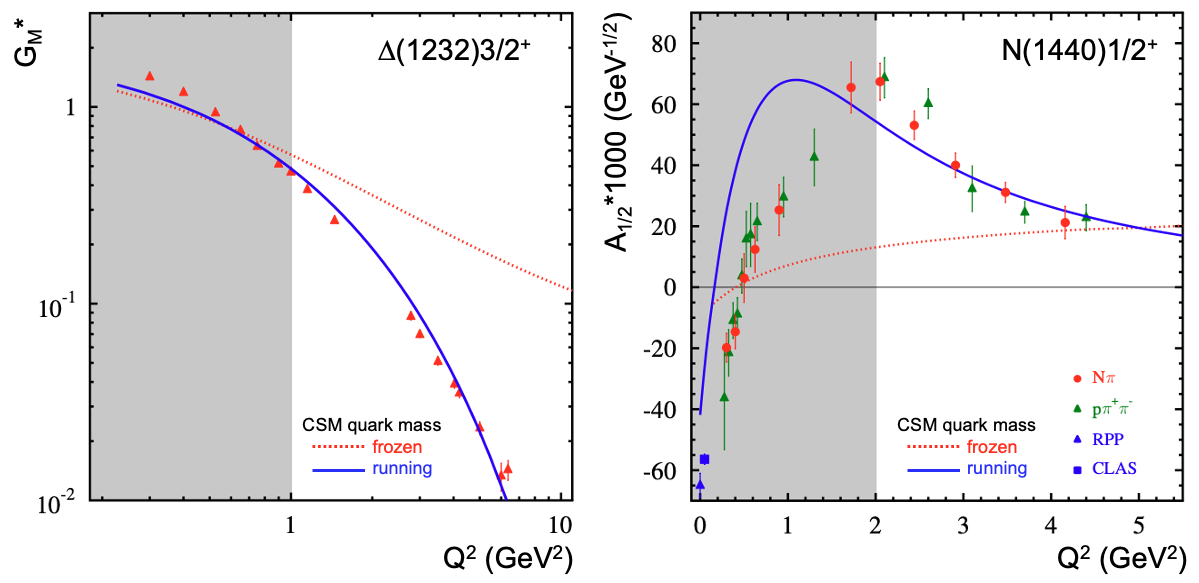}
\caption{Description of the $N \to \Delta$ magnetic transition form factor normalized to the dipole fit $G^\ast_M/3G_D$ with $G_D(Q^2)=(1+Q^2/0.71\,{\rm GeV}^2)^{-2}$ (left) and the electrocoupling $A_{1/2}$ for the $N(1440)1/2^+$ (right) achieved using CSMs~\cite{Wilson:2011aa, Segovia:2014aza, Segovia:2015hra}. 
Results obtained with a momentum-independent (frozen) dressed-quark mass~\cite{Wilson:2011aa, Segovia:2013uga} (dotted red curves) are compared with QCD-kindred results (solid blue curves) obtained with a momentum-dependent quark mass function of the type drawn in Fig.\,\ref{Quark_Gluon_massfunct}~\cite{Segovia:2014aza, Segovia:2015hra}. 
The electrocoupling data were taken from Refs.\,\cite{CLAS:2009ces, Villano:2009sn, CLAS:2014fml} for $\pi N$ electroproduction and Refs.\,\cite{Mokeev:2015lda, CLAS:2012wxw, Mokeev:2023zhq} for $\pi^+\pi^-p$ electroproduction. 
The photocouplings for the $N(1440)1/2^+$ are from the PDG 
\cite{PhysRevD.110.030001} and from Ref.~\cite{CLAS:2009tyz}, blue square and triangle, respectively. 
The approximate range of $Q^2$ where the contributions from the meson--baryon cloud remain substantial are highlighted in gray.
\label{csm_delta_roper}}
\end{figure}

CSM analyses of the $Q^2$ evolution of the electrocouplings of the $\Delta(1232)3/2^+$ and \linebreak $N(1440)1/2^+$ derived from experiments during the 6-GeV era at JLab have demonstrated the ability to gain insight into the momentum dependence of the dressed-quark mass function. 
As representative examples, the CSM predictions for the leading $N \to \Delta$ magnetic transition form factor in the Ash convention~\cite{Ash:1967rlw}, normalized to the dipole fit, and the $A_{1/2}$ electrocoupling for the $N(1440)1/2^+$ are shown in Fig.\,\ref{csm_delta_roper}. 
All electroexcitation amplitudes for the $\Delta(1232)3/2^+$ and $N(1440)1/2^+$ have been predicted within the CSM framework and can be found in Refs.\,\cite{Segovia:2013uga, Segovia:2014aza, Segovia:2015hra, Burkert:2017djo}. 
To study the impact of the dynamically generated momentum-dependent dressed quark mass, the electroexcitation amplitudes were obtained using: 
(\textit{a}) a simplified momentum-independent $qq$ contact interaction~\cite{Wilson:2011aa, Segovia:2013uga} and 
(\textit{b}) Schwinger functions that express momentum dependence of the sort driven by a $qq$-interaction derived from the QCD Lagrangian~\cite{Segovia:2014aza, Segovia:2015hra}.

In the simpler version (\textit{a}), the interaction kernel for the gap equation is treated within the rainbow-ladder (RL) truncation, which is leading-order in a symmetry-preserving and systematically improvable approximation scheme~\cite{Munczek:1994zz, Bender:1996bb}, and the interaction between quarks is expressed by a momentum-independent ``gluon'' propagator: 
\begin{equation}
\label{njlgluon}
g^2 D_{\mu \nu}(p-q) = \delta_{\mu \nu} \frac{4 \pi \alpha_{\rm IR}}{m_G^2}\,.
\end{equation}
Here, making connection with Eq.\,\eqref{RGIGluonMass}, $m_G=0.5\,$GeV, and the interaction strength is a parameter, with $\alpha_{\rm IR}/\pi=0.36$ providing for a good description of light-meson properties; see, {\it e.g.}, Refs.\,\cite{Gutierrez-Guerrero:2010waf, Chen:2012txa}.  
Version (\textit{a}) defines an effective field theory --- a symmetry-preserving treatment of a contact interaction (SCI) 
\cite{Gutierrez-Guerrero:2010waf, Chen:2012txa} --- designed to capture the near-conformal behavior of QCD's process-independent effective charge, \textit{viz}., the fact that $\hat\alpha$ does not run on $k\lesssim m_0/2$; see Fig.\,\ref{Quark_Gluon_massfunct}\,-\,right.
The SCI is not a precision tool, but it does have many merits.
For instance, 
algebraic simplicity;
simultaneous applicability to a diverse array of systems and processes;
potential for revealing insights that link and explain numerous phenomena;
and capability to serve as a tool for checking the validity of algorithms employed in computations that depend upon high performance computing.
Significantly, today's applications are typically parameter-free.

Using the SCI, one predicts a momentum-independent dressed-quark mass $M\approx 0.35\,$GeV, in line with the far infrared behavior of the running quark mass in Fig.\,\ref{Quark_Gluon_massfunct}\,-\,left.
One use of SCI analyses is that they expose those features of $N\to N^\ast$ transition form factors that are (most) sensitive to QCD-like running of propagators and vertices.  
As examples, SCI predictions for the magnetic transition form factor of the $\Delta(1232)3/2^+$, normalized to the dipole fit $G_M^\ast/3G_D$, and the $A_{1/2}$ electrocoupling for the $N(1440)1/2^+$ are shown in Fig.~\ref{csm_delta_roper} as red dashed lines~\cite{Wilson:2011aa, Segovia:2013uga}.
The calculations overestimate the experimental results for the $\Delta(1232)3/2^+$ on $Q^2 > 1\,$GeV$^2$, with the discrepancy increasing with $Q^2$, reaching more than an order of magnitude at $Q^2 > 4\,$GeV$^2$. 

As seen in Fig.\,\ref{csm_delta_roper}\,-\,right, apart from producing a zero crossing in the correct neighborhood, the SCI fails to describe the $N(1440)1/2^+$ electrocouplings across the entire range of $Q^2$ covered by the measurements. 
Plainly, as intended by the calculations, one can see that experimental results on the $Q^2$ evolution of the $\Delta(1232)3/2^+$ and $N(1440)1/2^+$ electrocouplings cannot be reproduced by a dynamically generated but frozen quark mass.

On the other hand, approach (\textit{b}), built upon Schwinger functions that express momentum dependence of the sort driven by QCD's $qq$-interaction~\cite{Segovia:2014aza, Segovia:2015hra}, delivers the solid blue curves in Fig.\,\ref{csm_delta_roper}.
In this case, a good description of the electroexcitation amplitudes for both resonances is achieved across the full range of $Q^2$ where the quark core contributions are expected to dominate. 

The $N^\ast$ electroexcitation experiment-theory comparisons in Fig.\,\ref{csm_delta_roper} provide strong support for the conclusion that {\it the mass of dressed quarks runs with distance, which may be defined by the inverse of the photon virtuality $Q^2$.} 
Evidence for the relevance of constituent quarks with running masses as the active component in $N^\ast$ structure has also been obtained from studies of the $\gamma_v pN^\ast$ electrocouplings in quark models~\cite{Aznauryan:2018okk, Aznauryan:2016wwm, Aznauryan:2012ec}.

Significantly, this good description of the JLab results on the electroexcitation amplitudes of the $\Delta(1232)3/2^+$ and $N(1440)1/2^+$ was achieved when using the same dressed quark mass function that is also employed in the successful description of pion and nucleon elastic electromagnetic form factors~\cite{Yao:2024drm, Yao:2024uej}. 
Consistent results on the dressed quark mass function, obtained from independent studies of pion and ground state nucleon structure and from the $N^\ast$ electroexcitation amplitudes for states with different structures --- spin--isospin flip for the $\Delta(1232)3/2^+$ and the first radial excitation of dressed quarks for the $N(1440)1/2^+$ --- provide compelling evidence for the existence of dynamically generated dressed quarks with running masses as the principal active components in the structure of pions, kaons, ground state nucleons, and the $\Delta(1232)3/2^+$ and $N(1440)1/2^+$. 
This success also demonstrates the capability of drawing a map of the RGI dressed quark mass function using experimental results on the $Q^2$ evolution of $N^\ast$ electroexcitation amplitudes when analyzed using CSMs.

\begin{figure}[t]
\begin{center}
\vspace{-2mm}
\includegraphics[width=0.98\textwidth]{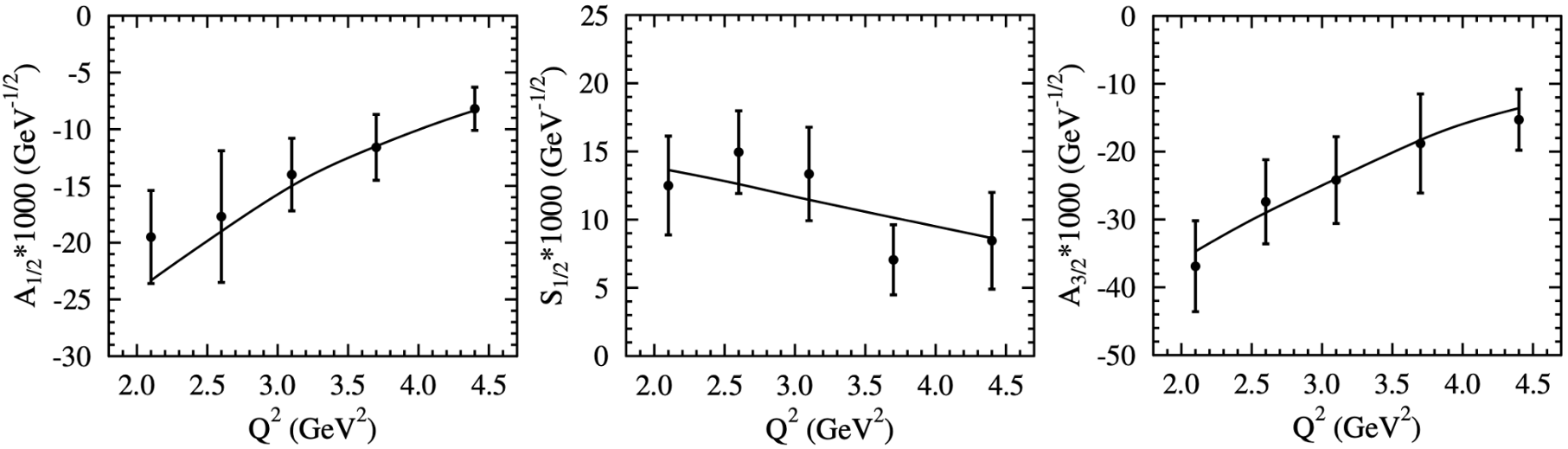}
\caption{$\Delta(1600)3/2^+$ electrocouplings obtained from the $\pi^+\pi^-p$ electroproduction cross sections measured with the CLAS detector~\cite{Mokeev:2023zhq}: $A_{1/2}$ (left), $S_{1/2}$ (center), and $A_{3/2}$ (right) in comparison with the CSM predictions~\cite{Lu:2019bjs}.}
\label{p33_csm_exp}
\end{center}
\end{figure}

\begin{figure*}[t]
\begin{center}
\includegraphics[width=0.85\textwidth]{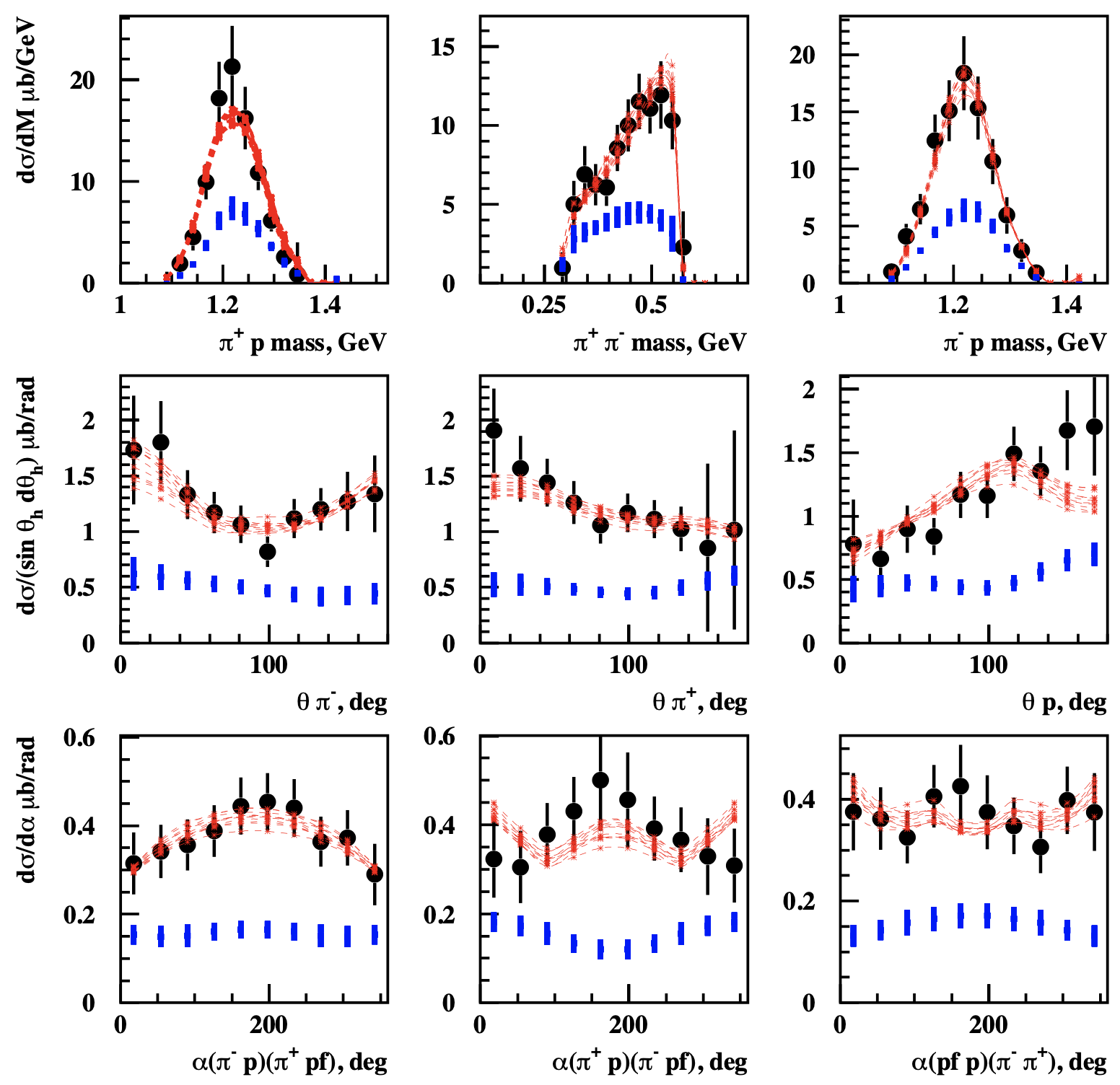}
\caption{Fits of the nine one-fold differential $\pi^+\pi^-p$ electroproduction cross sections measured with CLAS~\cite{Trivedi:2018rgo, CLAS:2017fja} (in black) achieved within the JM23 model~\cite{Mokeev:2023zhq} for $W$ from 1.500 -- 1.525~GeV and $Q^2$ from 3.0 -- 3.5~GeV$^2$. 
The data point uncertainties are evaluated as a quadratic sum of the statistical and relevant systematic uncertainties. 
The groups of red curves represent the JM23 fits closest to the data. 
The resonant contributions are shown in blue.} 
\label{full_nstar_146151156_31}
\end{center}
\end{figure*}

CSM predictions for the electrocouplings of the $\Delta(1600)3/2^+$ were made in 2019~\cite{Lu:2019bjs}, prior to the availability of experimental results.
These predictions are represented by the black solid lines in Fig.\,\ref{p33_csm_exp} compared to the available data from CLAS~\cite{Mokeev:2023zhq}. 
The first results on the electrocouplings of the $\Delta(1600)3/2^+$ were obtained in 2023 through the analysis of the nine one-fold differential $\pi^+\pi^-p$ cross sections in bins of $W$ and $Q^2$. 
The analysis covered a $W$-range from 1.46 -- 1.66\,GeV and $Q^2$ from 2.0\,GeV$^2$ to 5.0\,GeV$^2$. 
A representative example of the analyzed nine one-fold differential cross sections for a particular $W$ and $Q^2$ bin is shown in Fig.\,\ref{full_nstar_146151156_31}. 
These cross sections consist of:
\begin{itemize}
 \item three invariant mass distributions for the three pairs of final state hadrons,  
 \item three CM distributions over polar $\theta$-angles for the final state $\pi^+$, $\pi^-$, and $p$, and  
 \item three distributions over the $\alpha$-angles between the two planes A and B defined by A) the three-momentum vectors of a pair of final state hadrons and B) of the third final hadron and the initial state photon in the CM frame as described in Ref.~\cite{Mokeev:2023zhq}.
\end{itemize}

The analysis was performed using the JM meson--baryon reaction model, developed for extracting the $\gamma_vpN^\ast$ photo- and electrocouplings by fitting the $\pi^+\pi^-p$ photo- and electroproduction observables~\cite{Mokeev:2023zhq, Mokeev:2015lda, CLAS:2012wxw, Mokeev:2008iw, Ripani:2000va}. 
The JM model achieves a good description of the $\pi^+\pi^-p$ differential cross sections over the range $W$ = 1.4 -- 2.0~GeV and $Q^2$ up to 5\,GeV$^2$. In fitting the CLAS $\pi^+\pi^-p$ electroproduction data~\cite{Trivedi:2018rgo,CLAS:2017fja}, the $\gamma_vpN^\ast$ electrocouplings 
were varied along with their total $\Gamma_{tot}$ and partial $\Gamma_{\pi \Delta}$, $\Gamma_{\rho p}$ decay widths to the $\pi \Delta$ and $\rho p$ final states, respectively, as well as their masses. In Ref.~\cite{Mokeev:2023zhq} the $\gamma_vpN^\ast$ electrocouplings were varied for only the resonances in the mass range up to 1.6\,GeV. This analysis was focused on extraction of the $\gamma_vpN^\ast$ electrocouplings for excited states of the nucleon located within the $W$ range from 1.4 -- 2.0~GeV. 
Additionally, the parameters of the non-resonant amplitudes described in Ref.\,\cite{Mokeev:2023zhq} were varied simultaneously.  

For each trial set of parameters, nine one-fold differential cross sections were computed using the JM23 model~\cite{Mokeev:2023zhq} and the $\chi^2/\text{data-point (d.p.)}$ was evaluated through a point-by-point comparison with the measured cross sections in each $W$ and $Q^2$ bin. 
The computed cross sections closest to the data, achieving $\chi^2/\text{d.p.}$ below a predetermined threshold (defined to ensure the computed cross sections lie within the data uncertainties for most measured points), were selected. 
These selected computed cross sections are shown in Fig.\,\ref{full_nstar_146151156_31} by the red curves, while the resonant contributions are shown by the blue bars with vertical sizes representing their uncertainties.  

A good description of the measured cross sections enables the reliable isolation of the resonant contributions. 
Notably, the uncertainties of the resonant contributions are comparable to those of the measured cross sections, confirming an unambiguous separation between the resonant/non-resonant contributions. 
In the JM23 model, the resonant amplitudes are parameterized within a unitarized Breit-Wigner \textit{Ansatz}~\cite{CLAS:2012wxw, Aitchison:1972ay}, ensuring consistency with the general unitarity condition for resonant amplitudes. 
From the fit, the resonance $\gamma_vpN^\ast$ electrocouplings, masses, and hadronic decay widths were extracted. 
The resonance parameters for the selected computed cross sections were averaged, with the mean values treated as the extracted results and the root-mean-square deviations of the parameters taken as their uncertainties.
       
The masses, total decay widths, and partial decay widths to the $\pi \Delta$ final state of the $\Delta(1600)3/2^+$, obtained from independent fits of the data in three overlapping $W$ intervals (1.46 -- 1.56~GeV, 1.51 -- 1.61~GeV, and 1.56 -- 1.66~GeV) for two $Q^2$ ranges (2.0 -- 3.5~GeV$^2$ and 3.0 -- 5.0~GeV$^2$), are presented in Table~\ref{hadr_d1600}. The extracted mass and decay widths of the $\Delta(1600)3/2^+$ were found to be $Q^2$-independent within their uncertainties.

This result provides strong evidence that the $\Delta(1600)3/2^+$ is excited in the $s$-channel for virtual photon--proton interactions, supporting the presence of a quark core, as predicted by the CSM~\cite{Lu:2019bjs}, and consistent with the description of this resonance in quark models~\cite{Aznauryan:2016wwm, Aznauryan:2015zta, Giannini:2015zia}. 
The findings from the analysis of $\pi^+\pi^-p$ electroproduction are inconsistent with the description of the $\Delta(1600)3/2^+$ as entirely dynamically generated by singularities in amplitudes other than resonance excitation in the $s$-channel~\cite{Leinweber:2024psf}, as the non-resonant $\pi^+\pi^-p$ electroproduction amplitudes show a pronounced dependence on $Q^2$, making it virtually impossible to describe the data across a broad $Q^2$ range using $Q^2$-independent masses and decay widths.

\begin{table*}
\begin{center}
\begin{tabular}{cccccc} 
\toprule
$W$ Interval    & $Q^2$ Interval   & Mass          & $\Gamma_{tot}$ & $\Gamma_{\pi\Delta}$ & $BF_{\pi\Delta}$ \\
(GeV)           & (GeV$^2)$        & (GeV)         & (MeV)          & (MeV)                & (\%)             \\ 
\midrule
1.46 -- 1.56     & 2.0 -- 3.5        & 1.55 $\pm$ 0.014 & 244 $\pm$ 21      &  154 $\pm$ 21          & 50 -- 78 \\ \midrule
1.51 -- 1.61     & 2.0 -- 3.5        & 1.57 $\pm$ 0.018 & 259 $\pm$ 21      &  169 $\pm$ 22          & 52 -- 81 \\ \midrule
1.56 -- 1.66     & 2.0 -- 3.5        & 1.57 $\pm$ 0.042 & 256 $\pm$ 33      &  166 $\pm$ 34          & 46 -- 90 \\ \midrule
1.46 -- 1.56     & 3.0 -- 5.0        & 1.56 $\pm$ 0.030 & 249 $\pm$ 37      &  158 $\pm$ 37          & 42 -- 92 \\ \midrule
1.51 -- 1.61     & 3.0 -- 5.0        & 1.56 $\pm$ 0.030 & 249 $\pm$ 34      &  158 $\pm$ 34          & 44 -- 89 \\ \midrule
1.56 -- 1.66     & 3.0 -- 5.0        & 1.58 $\pm$ 0.039 & 263 $\pm$ 29      &  172 $\pm$ 29          & 49 -- 86 \\ \midrule
\multicolumn{2}{c}{PDG}      & 1.50 -- 1.64      & 200 -- 300         &                      & 58 -- 82 \\ 
\bottomrule
\end{tabular}
\caption{Mass and total/partial decay widths of the $\Delta(1600)3/2^+$ to the $\pi\Delta$ final state determined from the fit~\cite{Mokeev:2023zhq} of the $\pi^+\pi^-p$
electroproduction cross sections~\cite{Trivedi:2018rgo, CLAS:2017fja} carried out independently within two intervals in $Q^2$ and within three overlapping intervals in 
$W$. The PDG~\cite{PhysRevD.110.030001} parameters are listed in the bottom row.
\label{hadr_d1600} }
\end{center}
\end{table*}

The $\Delta(1600)3/2^+$ contributes significantly to the $\pi^+\pi^-p$ electroproduction amplitudes across all three $W$-intervals analyzed. 
The non-resonant amplitudes differ among these intervals. 
Consistent results for the $\Delta(1600)3/2^+$ electrocouplings across the three $W$-intervals emphasize the credibility of these extractions. 
The values of the electrocouplings averaged over the three $W$ intervals by employing the procedure described in Section IV\,A of Ref.\,\cite{Mokeev:2023zhq} are shown in Fig.\,\ref{p33_csm_exp} by the data points with uncertainties. 
Note, the uncertainties account both for experimental data statistical and bin-by-bin systematic uncertainties, and for the uncertainties of the $N^\ast$ hadronic decay widths, masses, and non-resonant amplitudes of the JM23 model. The experimental results for the $\Delta(1600)3/2^+$ electrocouplings confirmed the CSM predictions made four years earlier~\cite{Lu:2019bjs}. 
This success further solidifies evidence for the capability to map the momentum dependence of the dressed quark mass using experimental results on the $Q^2$ evolution of the $\gamma_vpN^\ast$ electrocouplings.

CSM analyses of experimental results on the $Q^2$ evolution of the $\gamma_vpN^\ast$ electrocouplings, combined with data on pion and nucleon elastic form factors, have provided overwhelming evidence for insights into the strong interaction dynamics underlying the emergence of hadron mass. 
The successful description of the aforementioned observables for meson and baryon structure, achieved using the same momentum-dependent features of the dressed quark mass, strongly supports the CSM concept of EHM. 
It also provides justification for the approximations made in relating the properties of the active degrees of freedom in hadron structure --- dressed quarks and gluons --- to the QCD Lagrangian. 

Currently, CSMs present the only approach that enables, within a common framework based on the QCD Lagrangian, the interpretation of observables for both meson and baryon structure~\cite{Yao:2024drm, Yao:2024uej, Wang:2024fjt, Yao:2024ixu, Lu:2022cjx}. 
Progress in other QFT approaches capable of relating hadron structure observables to the QCD Lagrangian will be important for advancing our understanding of how the dominant part of hadron mass and structure are generated by strong interactions in the sQCD\,$\leftrightarrow$\,pQCD transition.

\begin{figure*}[t]
\begin{center}
\includegraphics[width=0.85\textwidth]{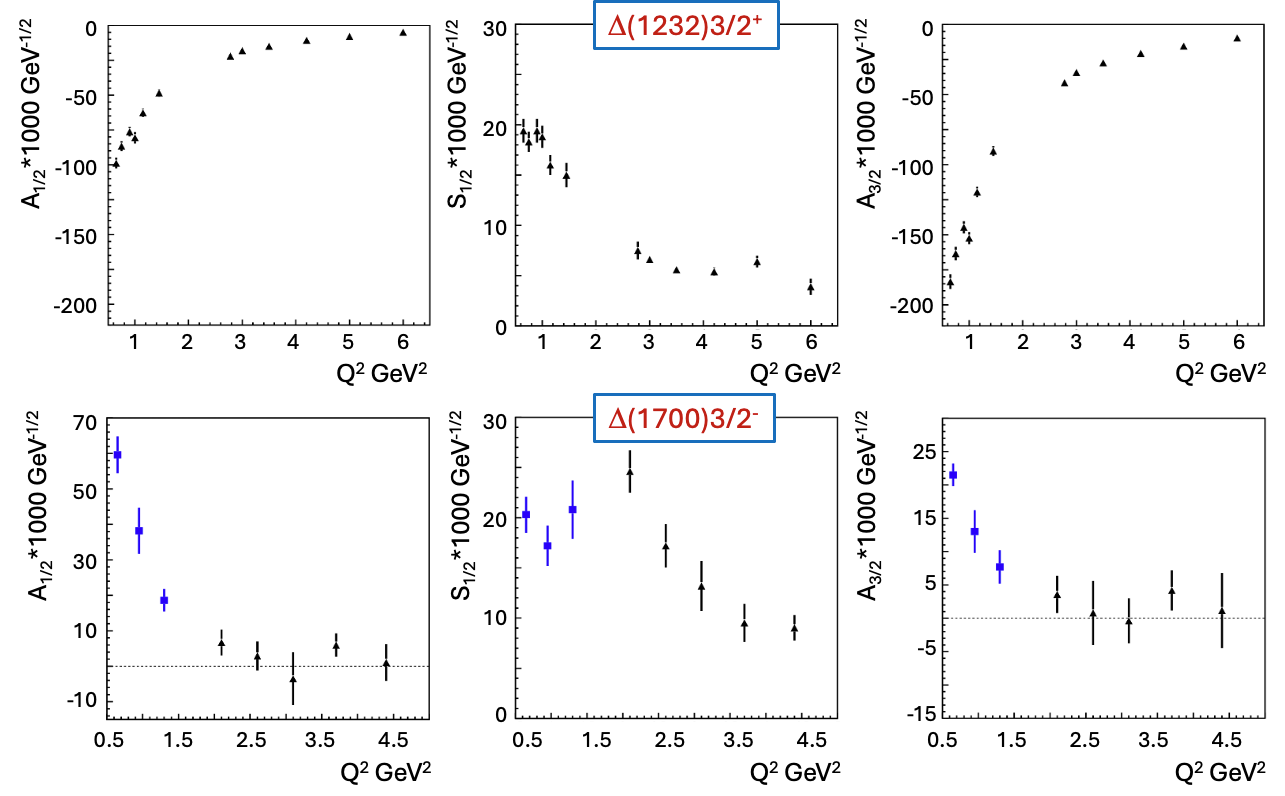}
\caption{(Top): Electrocouplings of the $\Delta(1232)3/2^+$ obtained from the analysis of $\pi N$ electroproduction data~\cite{CLAS:2009ces}. 
(Bottom): Electrocouplings of the $\Delta(1700)3/2^-$ obtained from the $\pi^+\pi^-p$ channel (blue) in combined studies of photo- and electroproduction~\cite{Mokeev:2020hhu} and (black) preliminary results for $Q^2$ from 2.0 to 5.0\,GeV$^2$~\cite{mokeev-nstar24}.} 
\label{d33_p33ground_el}
\end{center}
\end{figure*}

Recently, preliminary results on the $\gamma_vpN^\ast$ electrocouplings of most resonances in the third resonance region were obtained from the analysis of $\pi^+\pi^-p$ electroproduction within the JM23 reaction model~\cite{mokeev-nstar24}. 
The electrocouplings of the $\Delta(1700)3/2^-$, compared to those of the $\Delta(1232)3/2^+$, are presented in Fig.\,\ref{d33_p33ground_el}. 
These two resonances are considered chiral partners; see, \textit{e.g}., the discussion in Ref.\,\cite{Liu:2022nku}.
In the chiral limit, their electrocouplings must be related, as predicted by chiral symmetry for massless bare quarks. 
Any departure seen in the experimental results from the expectations in the chiral limit provides valuable insights into the role of DCSB in the generation of $N^\ast$ masses and their structure. 
The experimental results on the $Q^2$ evolution of the $\Delta(1223)3/2^+$ and $\Delta(1700)3/2^-$ electrocouplings reveal pronounced differences. 
The electroexcitation of the $\Delta(1223)3/2^+$ is dominated by the transverse amplitudes ($A_{1/2}$ and $A_{3/2}$) across the entire range of $Q^2$ covered by the data. 
In contrast, for the $\Delta(1700)3/2^-$, the longitudinal $S_{1/2}$ amplitude becomes dominant for $Q^2 > 2$\,GeV$^2$, while the transverse amplitudes are nearly zero.

CSMs express DCSB as a corollary of EHM. 
Future comparative CSM studies of the electrocouplings of the $\Delta(1232)3/2^+$ and $\Delta(1700)3/2^-$, particularly predictions of the $\Delta(1700)3/2^-$ electrocouplings for $Q^2$ up to 5\,GeV$^2$, will provide deeper insights into the relationship between EHM and DCSB.

Quark models describe the $\Delta(1700)3/2^-$ as a bound system of three quarks, with orbital momentum $L=1$ as the biggest component in its wavefunction~\cite{Giannini:2015zia}. 
In these quark models, this state is considered a member of the $[70,1^-]$ SU(6)\(\times\)O(3) supermultiplet. 
CSMs predict a far richer structure \cite{Liu:2022nku} and the CSM predictions for the $\gamma_vpN^\ast$ electrocouplings of other resonances in this supermultiplet, particularly for the $N(1520)3/2^-$ and $N(1535)1/2^-$, represent an important direction for future investigations. 
So far, CSMs have provided predictions for the \(\gamma_vpN^\ast\) electrocouplings of nucleon resonances that can be described as spin-isospin flips or first radial excitations of the quark-diquark system. 
Extending CSM predictions to include the electrocouplings of resonances with quarks in orbital excitations offers a promising opportunity to either establish the universality of the dressed quark mass function or reveal its sensitivity to the structural environment.

\section{Extending Insight into EHM from CLAS12 Experiments and Beyond}
\label{12gev22gev_baryons}
%
%
The domain of sQCD\,$\leftrightarrow$\,pQCD transition was discussed in connection with Eqs.\,\eqref{QCDbeta}\,--\,\eqref{QCDM0}.  
Therewith, it was argued that for $k\gtrsim 5m_0$ one is working with parton-like quarks, whose active mass can largely be attributed to Higgs boson couplings into QCD. 
By contrast, on the domain $k\lesssim 2 m_0$, Eq.\,\eqref{QCDM0}, quasiparticle quarks have emerged as the dominant valence degrees of freedom.  
These dressed quarks carry a fully emergent mass: $M \sim 0.4\,$GeV.
The behavior and evolution  of $M(k)$ on this domain is essentially nonperturbative; hence, definitive of EHM.  
Consequently, empirical validation of the EHM paradigm as the solution to the problem defined by QCD requires that a map be drawn of $M(k)$, from the far infrared to $k\simeq 5 m_0$.

Both the capability to probe specific distance scales in studies of the $\gamma_v pN^\ast$ electrocouplings and the focus of such probing is determined by $Q^2$. Assuming roughly equal sharing of $Q^2$ among the three dressed quarks, a notion justified by the character of bound-state wavefunctions, the accessible range of quark momenta for mapping the dressed quark mass function can be estimated as:
\begin{equation}
\label{krange}
k \approx \frac{\sqrt{Q^2}}{3}\,.
\end{equation}

\begin{figure*}[t]
\begin{center}
\includegraphics[width=0.85\columnwidth]{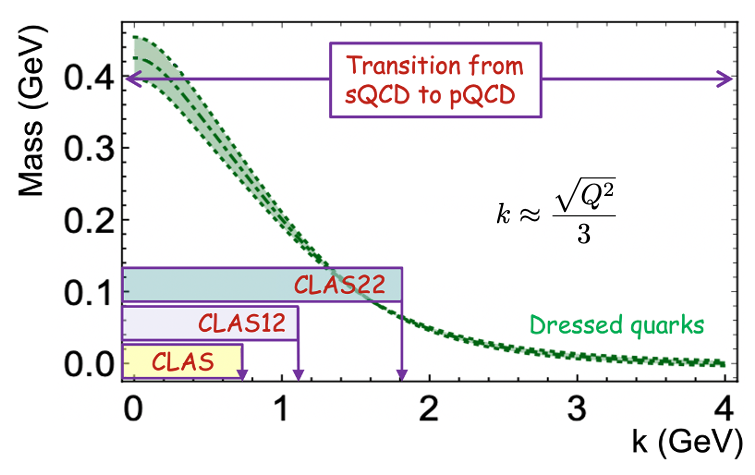}
\caption{Capabilities for mapping the momentum dependence of the dressed quark mass in studies of the $\gamma_v pN^\ast$ electrocouplings using data from CLAS, expected results from CLAS12, and projections for a potential CEBAF energy upgrade to 22~GeV, are presented in terms of the accessible ranges of quark momenta $k$.} 
\label{kranges61222}
\end{center}
\end{figure*}

JLab 6-GeV era experiments provided $Q^2$ coverage up to 5\,GeV$^2$ for most excited nucleon states with masses below 1.75\,GeV~\cite{Mokeev:2022xfo, Mokeev:2024beb}. In the near future the mass coverage will be extended up to 2.0\,GeV for $Q^2$ below 5\,GeV$^2$~\cite{mokeev-nstar24}.
The corresponding range of quark momenta accessible from these results for mapping the dressed quark mass function is drawn in Fig.\,\ref{kranges61222}.

The $\gamma_v pN^\ast$ electrocouplings extracted from CLAS measurements at $Q^2 < 5$\,GeV$^2$ allow exploration of the quark momentum range $k$ where less than 30\% of the dressed quark mass is generated. 
The portion of the emergent quark mass was evaluated as the difference between the fully dressed quark mass and the running quark mass at the maximal achievable value of $k$ from Eq.\,\eqref{krange}, corresponding to the maximum $Q^2$ reached in the measurements, normalized to the fully dressed quark mass. To extend the covered range of $k$, the $\gamma_v pN^\ast$ electrocouplings need to be obtained over a broader $Q^2$ range.

\begin{figure}[t]
\centering
\includegraphics[width=1.0\textwidth]{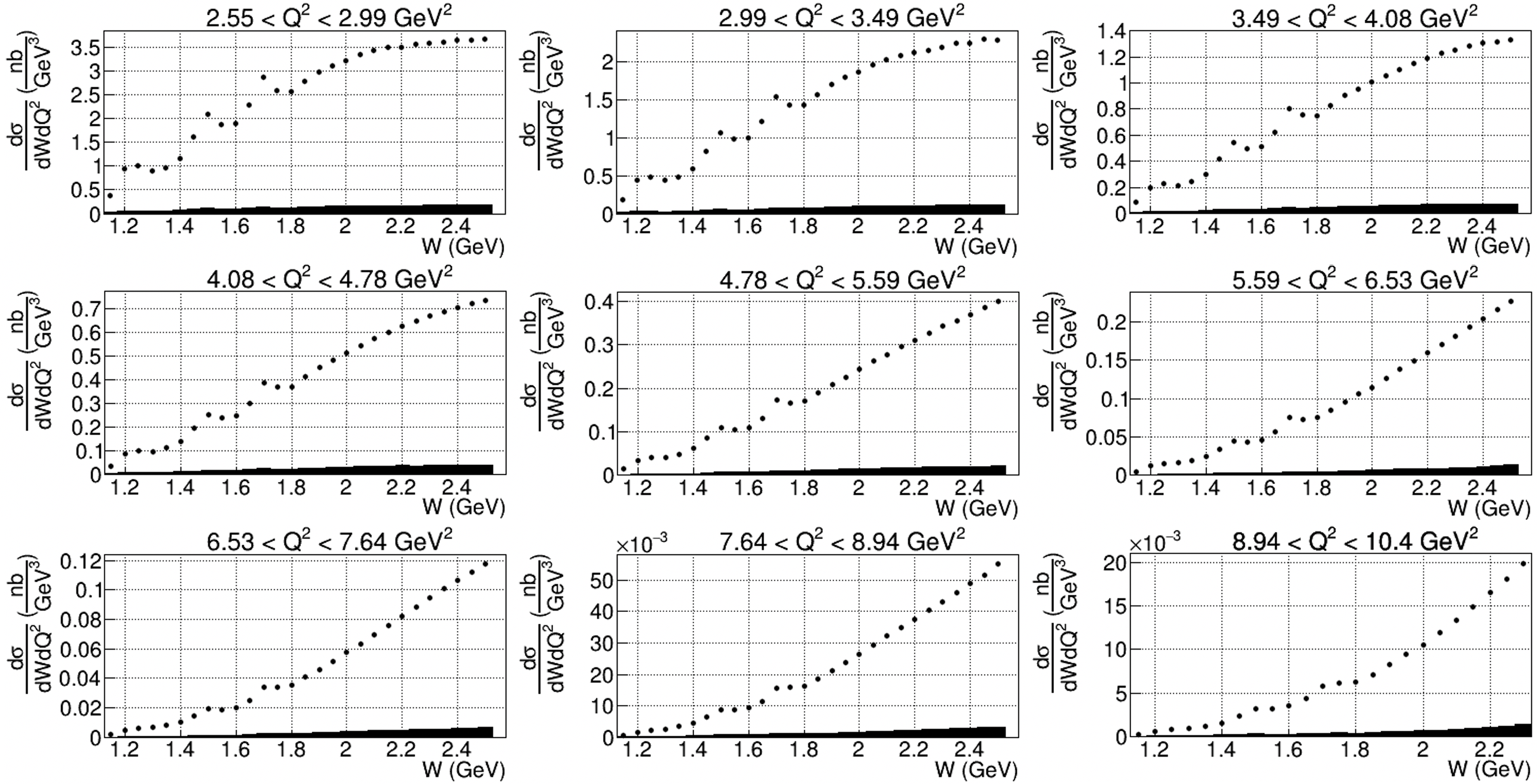}
\caption{Inclusive electron scattering cross sections determined from CLAS12 data in the nucleon resonance region for $Q^2$ from 2.55 -- 10.4\,GeV$^2$ \cite{CLAS:2025zup}. 
The statistical uncertainties are shown but they are smaller than the data point size for the majority of the data points. 
The bin-by-bin systematic uncertainties are shown by the filled area at the bottom of each plot.
\label{fig:XSEC_res}}
\end{figure}

\begin{figure}[t]
\centering
\includegraphics[width=0.98\textwidth]{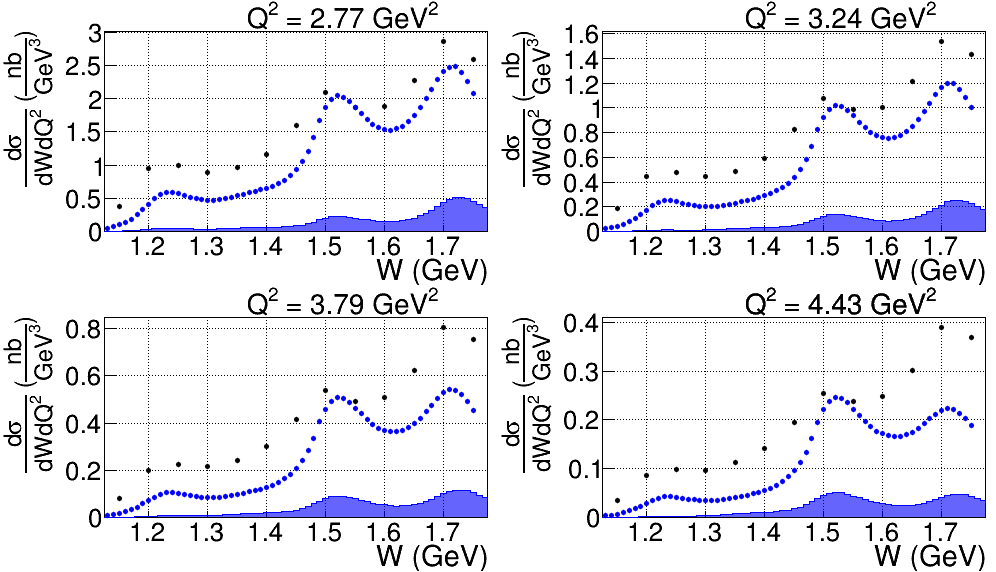}
\caption{Inclusive electron scattering cross sections from CLAS12 data (black points) as a function of $W$ for selected bins in $Q^2$ as shown. 
The blue points represent the computed resonant contributions from  experimental results on the resonance electrocouplings~\cite{HillerBlin:2019jgp, Blin:2021twt}. 
The shaded areas at the bottom of each plot show the systematic uncertainties associated with evaluation of the resonant contributions.
\label{incl-cs}}
\end{figure}

The ongoing measurements with CLAS12~\cite{achenbach-nstar24} provide a promising opportunity to extend results on the $Q^2$ evolution of the $\gamma_v pN^\ast$ electrocouplings into the hitherto unexplored range $Q^2 > 5$\,GeV$^2$. 
The first data on inclusive $(e,e'X)$ cross sections in the resonance region for $Q^2 < 10$\,GeV$^2$ have already become available from CLAS12~\cite{CLAS:2025zup} and are presented in Fig.\,\ref{fig:XSEC_res}. 
The structures in the $W$ dependence of the cross sections are generated by resonant contributions. 
This is clearly illustrated in Fig.\,\ref{incl-cs}, where the resonant contributions, evaluated from CLAS results on the $\gamma_v pN^\ast$ electrocouplings~\cite{HillerBlin:2019jgp, Blin:2021twt}, are shown. 
This observation highlights the potential to determine the $\gamma_v pN^\ast$ electrocouplings for $Q^2$ up to 10\,GeV$^2$ from measurements of exclusive meson electroproduction channels using CLAS12.

\begin{figure}[th]
\centering
\includegraphics[width=0.98\columnwidth]{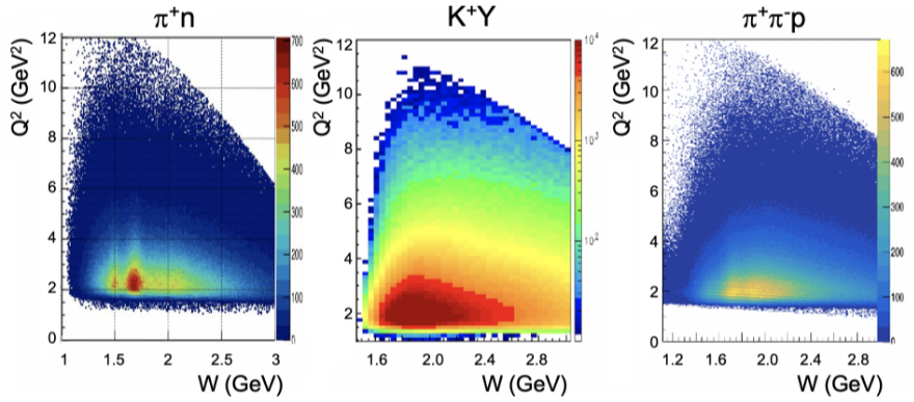}
\caption{The kinematic coverage $Q^2$ vs.\ $W$ for exclusive $\pi^+n$ (left), $K^+Y$ (Y=$\Lambda$, $\Sigma$) (center), and $\pi^+\pi^-p$ (right) electroproduction channels measured with CLAS12~\cite{mokeev-nstar24}.
\label{12GeV_excl}}
\end{figure}

CLAS12 is the only facility in the world capable of exploring exclusive meson electroproduction channels at photon virtualities reaching up to 10 -- 12\,GeV$^2$~\cite{achenbach-nstar24, mokeev-nstar24, carman-nstar24}.
These are the highest photon virtualities ever achieved in the resonance region of $W < 2.5$~GeV.
Figure~\ref{12GeV_excl} presents the kinematic coverage achieved in measurements with the CLAS12 detector for several exclusive meson electroproduction channels.

From the data on $\pi N$, $K\Lambda$, $K\Sigma$, and $\pi^+\pi^-p$ electroproduction, the $\gamma_v pN^\ast$ electrocouplings will be determined for most excited states of the proton in the mass range up to 2.5~GeV. 
These electrocouplings will be obtained through independent analyses of these exclusive channels, as well as from global multi-channel analyses using coupled-channels approaches~\cite{Wang:2024byt, Mai:2021aui, Matsuyama:2006rp, Burkert:2004sk, Suzuki:2010yn} updated for the description of exclusive electroproduction processes at $Q^2 > 5$\,GeV$^2$. 
Consistent results on the $\gamma_v pN^\ast$ electrocouplings from different exclusive channels and coupled-channels analyses will validate, in a nearly model-independent way, the credible extraction of these quantities. 
Extending amplitude analysis approaches to $Q^2$ up to 12\,GeV$^2$ is crucial for extracting reliable results on the $\gamma_v pN^\ast$ electrocouplings over this range. 
Analysis of these results within the CSM framework will enable mapping of the momentum dependence of the dressed quark mass across the range of quark momenta where approximately 50\% of the emergent mass component is generated; see Fig.\,\ref{kranges61222}. 
Furthermore, the results on the $\gamma_vpN^\ast$ electrocouplings for $Q^2 < 10$\,GeV$^2$ will allow us to map the range of distances where the sQCD\,$\leftrightarrow$\,pQCD transition in the QCD running coupling, $\alpha_s$, is expected; see Fig.~\ref{Quark_Gluon_massfunct}\,-\, right.

\begin{figure}[t]
\centering
\includegraphics[width=0.98\columnwidth]{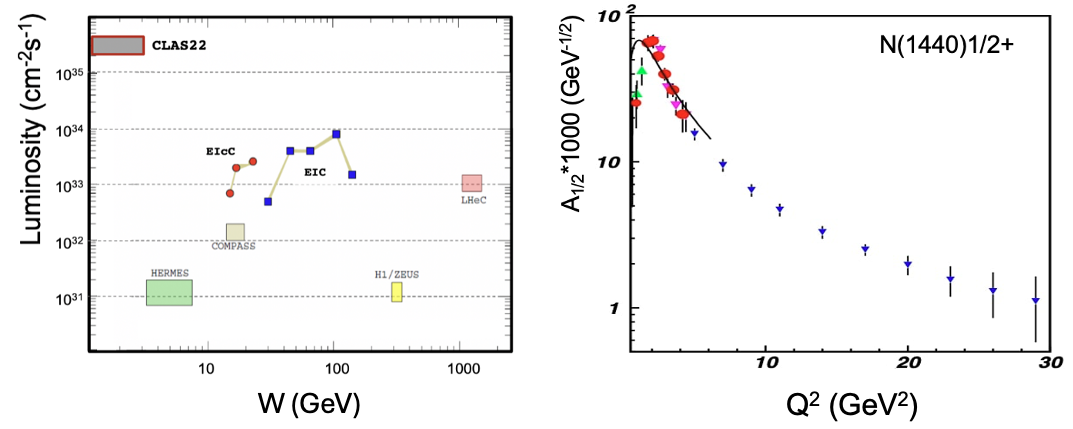}
\caption{(Left) Comparison of previous, current, and planned facilities for studying hadron structure in electroproduction, shown as a correlation plot of luminosity vs.\ $W$. (Right) Available (red and green) and projected results (blue) on the $Q^2$ evolution of the $N(1440)1/2^+$ electrocouplings~\cite{Accardi:2023chb}.}
\label{facilities}
\end{figure}

The development of the research program motivating the potential increase of CEBAF's energy to 22~GeV is currently underway~\cite{Accardi:2023chb}. 
Probing the momentum dependence of the dressed quark mass from experimental results on the $\gamma_v pN^\ast$ electrocouplings represents an important aspect of these efforts; see, {\it e.g.}, Section 6.5 of Ref.\,\cite{Accardi:2023chb}. Simulations of the $\pi N$, $K\Lambda$, $K\Sigma$, and $\pi^+\pi^-p$ electroproduction channels for an incoming electron beam energy of 22\,GeV with the configuration of the current CLAS12 detector~\cite{gothe-nstar24} have demonstrated the potential to extend information on the evolution of the $\gamma_v pN^\ast$ electrocouplings up to $Q^2 \approx 30$\,GeV$^2$. 
These extractions will be feasible if the large acceptance detection of the final-state products can also be achieved at increased luminosities, \textit{viz}., as high as $5 \times 10^{35}$\,cm$^{-2}$s$^{-1}$. 
A comparison of the luminosities of current and planned facilities for studying hadron structure in electroproduction experiments is shown in Fig.\,\ref{facilities}\,-\,left as a correlation plot of luminosity versus $W$.

None of the planned facilities, except CEBAF at 22\,GeV, will be capable of extending the results on the $Q^2$ evolution of the $\gamma_v pN^\ast$ electrocouplings into the $Q^2$ range from 10 -- 30\,GeV$^2$. 
The expected maximum luminosities for the EiC and EIcC electron--ion collider facilities in the USA and China are more than an order of magnitude below the levels required for such extractions. 
Additionally, the $W$ coverages provided by the EiC and EIcC do not support the exploration of nucleon resonance electroexcitations.

Only experiments with CEBAF at 22\,GeV, using an electron beam on a fixed target, will provide the necessary luminosity and $W$ coverage --- represented as CLAS22 in Fig.\,\ref{facilities}\,-\,left --- for extracting the $\gamma_v pN^\ast$ electrocouplings within the $Q^2$ range of 10 -- 30\,GeV$^2$. 
The available and projected results on the $Q^2$ evolution of $N(1440)1/2^+$ electrocouplings are shown in Fig.\,\ref{facilities}\,-\,right. Results of comparable quality are expected for all prominent resonances excited off protons.

Analyzed using CSMs, these results will offer the only foreseeable opportunity to explore the full range of distances where the dominant part of hadron mass and $N^\ast$ structure emerge from QCD. 
This makes CEBAF at 22\,GeV unique and potentially the ultimate QCD facility at the luminosity frontier.

\section{EHM from Combined Studies of Meson and Baryon Structure}
\label{mb_structure}
Studies of the structure of pseudoscalar mesons in concert with studies of the structure of the ground and excited states of the nucleon are crucial for understanding EHM. 
The SU(3) octet of light pseudoscalar mesons, consisting of pions, kaons, and the $\eta$-meson, represents the lightest bound $q\bar{q}$ systems in Nature. Additionally, the three pions and four kaons are the Goldstone bosons that emerge as a consequence of DCSB, \textit{viz}., they are massless in the chiral limit. 
(The $\eta$-meson and its partner, $\eta'$, are special because of the non-Abelian anomaly~\cite{Christos:1984tu, Ding:2018xwy}.)

Pseudoscalar mesons are exceptionally light compared to the typical hadronic mass scale. For instance, the pion mass is similar to that of the $\mu$-lepton, much smaller than the mass of the $\rho$ meson, and smaller than the sum of the fully dressed quark masses used in constituent quark models by a factor of six. Any comprehensive concept of EHM must describe not only the mass generation of baryons but also explain why pseudoscalar mesons are so light.  

Currently, CSMs represent the only QCD-connected approach capable of simultaneously describing both meson and baryon structure within a single unified framework; see, {\it e.g.}, Refs.\,\cite{Yao:2024drm, Yao:2024uej, Yao:2024ixu}.
Combined CSM studies of pseudoscalar mesons and ground and excited state nucleons address the following key challenges:
\begin{itemize}
    \item {\bf Unified framework for the emergence of meson and baryon mass and structure:} Assess the feasibility of understanding EHM for both mesons and baryons in relation to the QCD Lagrangian.
    \item {\bf Small masses of pseudoscalar mesons:} Explain the small masses of pseudoscalar mesons in connection with their dual nature as $q\bar q$
    bound systems and as Goldstone bosons.
    \item {\bf Understanding DCSB:} Extend insights into DCSB through studies of the dressed quark mass function in pions, kaons, and nucleon ground and excited states.
    \item {\bf Interplay between Higgs-generated and emergent parts of hadron mass and structure:} Elucidate the interplay between the Higgs and emergent mechanisms in hadron mass generation and interactions.
\end{itemize}

The SM remains incomplete until verified explanations account for the emergence of the nucleon ground and excited state masses in the 1 -- 3~GeV range, the significantly lower pion and kaon masses in the few hundred MeV range, and their associated phenomena. 
Additionally, the modulations of these effects by interactions with the Higgs boson are crucial to understand the formation and evolution of our Universe.

DCSB is empirically revealed most clearly through the properties of the pion. In fact, as discussed in Ref.\,\cite{Roberts:2016vyn}, in the chiral limit a set of Goldberger--Treiman--like relations~\cite{Maris:1997tm, Qin:2014vya, Binosi:2016rxz} establishes a connection between the pion's Bethe--Salpeter amplitude and the scalar component of the dressed-quark self-energy, $B(k^2)$. The most well-known of these relations states:  
\begin{equation}
\label{GT_eqs}
f_\pi E_\pi(k;P=0) = B(k^2), 
\end{equation}
where $E_\pi$ is the leading component of the pion's Bethe--Salpeter amplitude and $f_\pi$ is the leptonic decay constant of the pion. 
$B(k^2)$ defines the running dressed-quark mass in Eq.\,\eqref{PropQuark} through the relation $M(k^2) = B(k^2) Z(k^2)$. 
$P$ is the pion or the total $q\bar{q}$ bound-state four-momentum and $k$ is the relative four-momentum between the $q$ and $\bar{q}$. 
This equation holds exactly in chiral QCD and highlights the fact that the Nambu--Goldstone theorem fundamentally expresses an equivalence between the quark one-body problem and the two-body bound-state problem in QCD's color-singlet, flavor-nonsinglet, pseudoscalar channel. 
Consequently, and quite strikingly, the properties of the nearly massless pion are the clearest manifestation of the mechanism responsible for (almost) all visible mass in the Universe.  

In the chiral limit, pions and kaons are massless Goldstone bosons, with identical structure expressed through common Bethe-Salpeter amplitudes. In actuality, chiral symmetry is explicitly broken by quark-Higgs couplings. Therefore, studies of pion and kaon structure in experiments with electromagnetic probes, when interpreted using CSMs, which are capable of accounting for the interplay between the Higgs and emergent mechanisms, offer a direct window onto the dressed-quark mass function, $M(k^2)$. 
This function can also be independently probed through study of the ground and excited states of the nucleon. 
A successful description of the elastic form factors of the pion, kaon, and nucleon, as well as the $\gamma_vpN^\ast$ electrocouplings, employing the same dressed-quark mass function, $M(k^2)$, is crucial to validate the relevance of dressed quarks with running mass as an active component that defines hadron structure.

\begin{figure}[!t]
  \centering
    \begin{tabular}{c}
        {\sf A} \\[-0.1ex]
        \includegraphics[clip, width=0.45\textwidth]{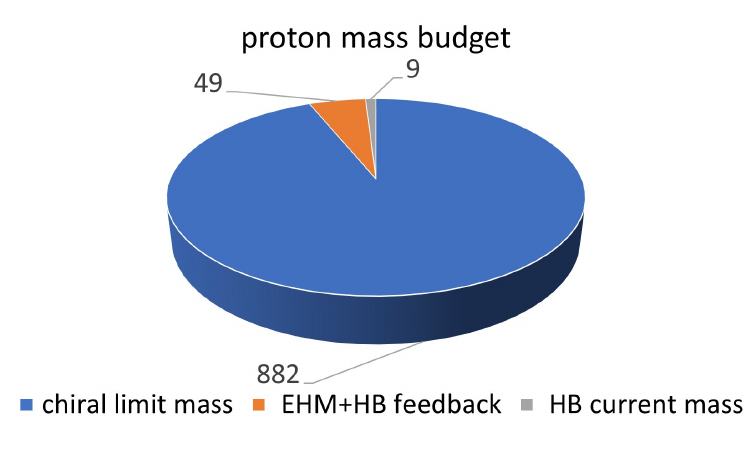}
    \end{tabular}
\hspace*{-8ex}\begin{tabular}{cc}
        {\sf B} & {\sf C} \\[-0.5ex]
        \includegraphics[clip, width=0.45\textwidth]{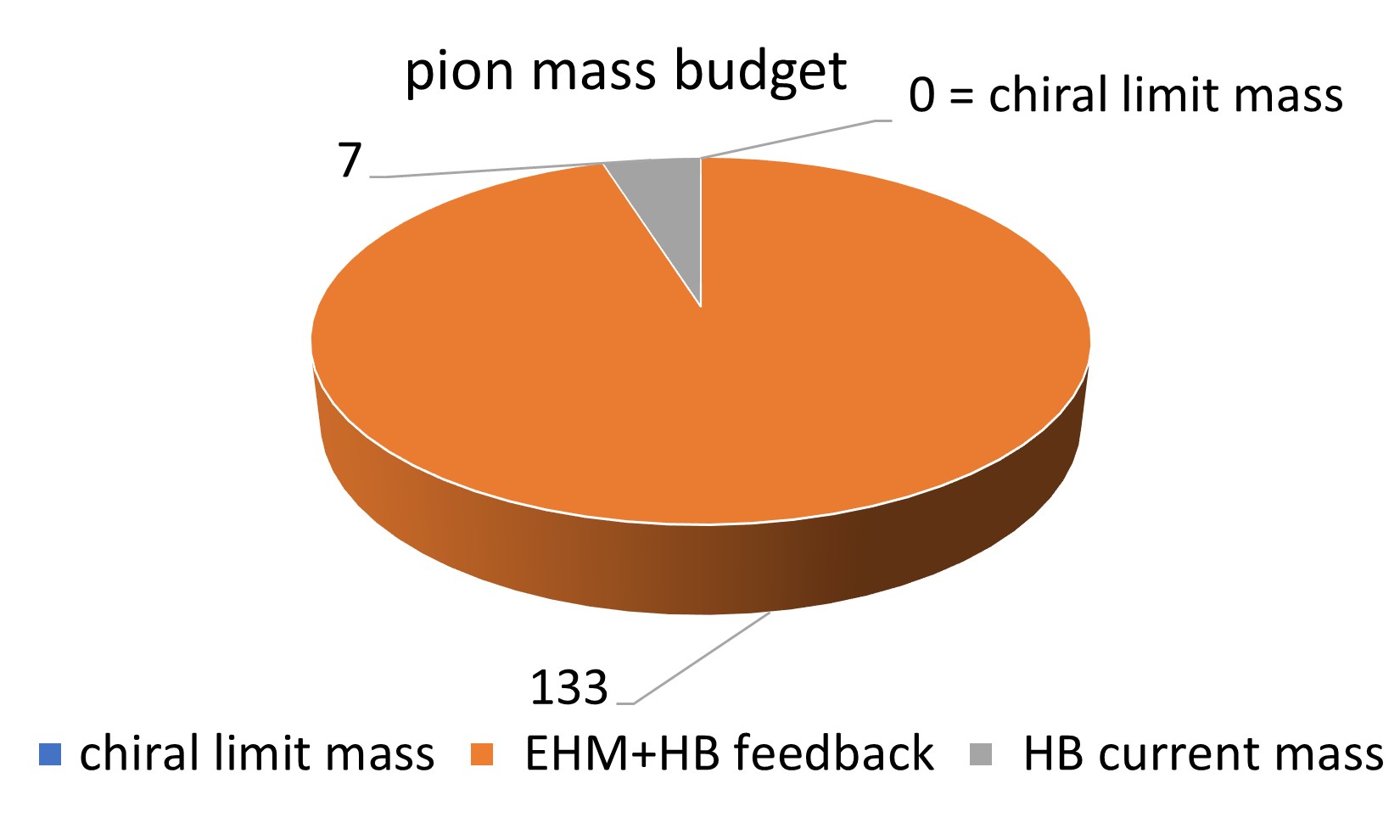} &
        \includegraphics[clip, width=0.45\textwidth]{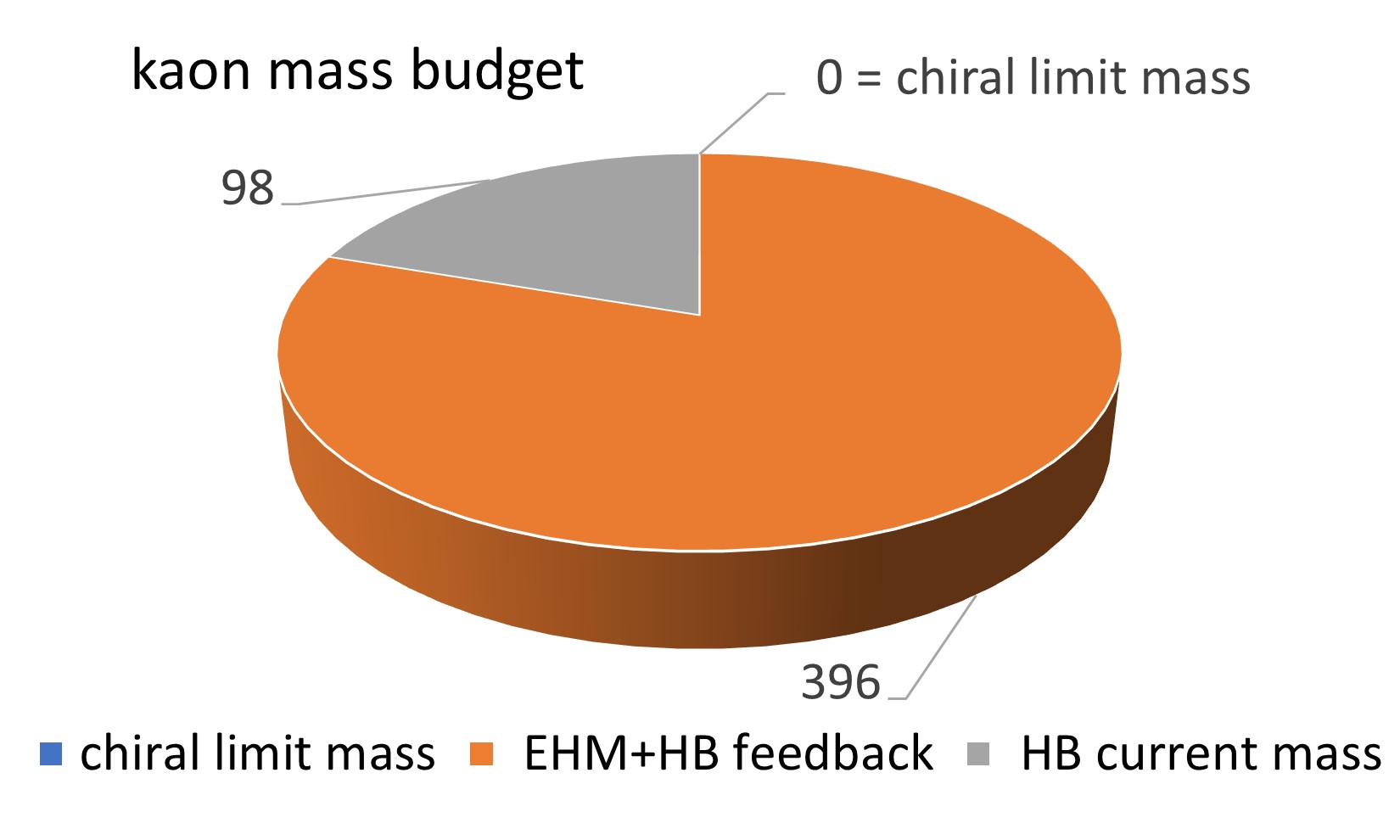}
    \end{tabular}
\caption{Mass budgets for {\sf A}\,--\,proton, {\sf B}\,--\,pion, and {\sf C}\,--\,kaon, drawn using a Poincar\'e invariant decomposition. Units MeV, separation at the renormalization scale of $\zeta=2$\,GeV - produced using information from Refs.\,\cite{Flambaum:2005kc, RuizdeElvira:2017stg, FlavourLatticeAveragingGroup:2019iem, PhysRevD.110.030001}.
\label{massbudgets}}
\end{figure}

The dual nature of pions and kaons as $q\bar{q}$ bound systems and Goldstone bosons is reflected in the distinctive features of their mass budget; see Fig.\,\ref{massbudgets}\,B, C.
The contributions from the emergent mass components of pions and kaons are explicitly zero in the chiral limit, yet at empirical quark current masses, these emergent components drive their masses to larger values through interference with the Higgs mechanism. 
Specifically, chiral symmetry imposes the following relation between the current masses of the valence quarks, $m_{f,g}^\zeta$, and the masses of the Nambu--Goldstone bosons (pions and kaons)~\cite{Maris:1997tm, Qin:2014vya}:
\begin{equation}
\label{MGMOR}
f_{NG} \, m_{NG}^2 = (m_f^\zeta + m_g^\zeta) \rho_{NG}^\zeta\,,
\end{equation}
where $f_{NG}$ is the meson's leptonic decay constant, {\it i.e.}, the pseudovector projection of the meson wavefunction onto the origin in configuration space, and $\rho_{NG}^\zeta$ is its pseudoscalar analogue.
For ground state pseudoscalar mesons, both $f_{NG}$ and $\rho_{NG}^\zeta$ are order parameters for chiral symmetry breaking.  

According to Eq.\,\eqref{MGMOR}, in the chiral limit, the masses of Goldstone bosons should be zero. 
With the Goldberger--Treiman-like relations, one can construct an algebraic proof~\cite{Maris:1997tm, Qin:2014vya, Binosi:2016rxz} showing that at any and every order within a symmetry-preserving truncation of the quantum field theory equations necessary to describe a pseudoscalar bound state, there is an exact cancellation between the mass-generating effect of dressing the valence quark and antiquark, which constitute the system, and the attractive interaction between them:
\begin{align}
M^{\rm dressed}_{\rm quark} + M^{\rm dressed}_{\rm antiquark} + U^{\rm dressed}_{\rm quark-antiquark\;interaction} \stackrel{\rm chiral\;limit}{\equiv} 0\,.
\label{EasyOne}
\end{align}
This is the mechanism that drives the seeming ``disappearance'' of the scale anomaly in the chiral-limit pion.  

An analogy with quantum mechanics emerges: the mass of a QCD bound state is the sum of the characteristic mass scales of its constituents plus (a negative, and sometimes substantial) binding energy. 
The mass budgets for the pion, kaon, and proton are shown in Fig.\,\ref{massbudgets}.

There are crucial differences in the mass budgets of the pion, kaon, and proton. The proton’s mass remains large even in the chiral limit, {\it i.e.}, even in the absence of Higgs boson (HB) couplings within QCD. This non-zero chiral-limit component reflects the contribution of emergent hadron mass in the SM. Indications are that the same is also qualitatively true for nucleon resonances.

When Higgs couplings for quarks inside the ground and excited proton states are considered, two additional contributions appear, as illustrated in Fig.\,\ref{massbudgets}\,A. 
Here the gray wedge represents the sum of the valence quark current masses, which contribute only $\sim$1\% to the total mass of the proton and $N^\ast$s. 
This wedge marks the sole Higgs-only component of a hadron's mass.
The remaining slice (orange) represents the mass fraction generated by constructive interference between EHM and HB effects. 
This contribution is largely determined by the in-proton expectation value of the chiral condensate operator~\cite{Brodsky:2010xf, Chang:2011mu, Brodsky:2012ku}:  
\begin{equation}
\langle p(P) |\bar{q}_{\mathpzc f} q_{\mathpzc f}| p(P) \rangle,
\end{equation}
\noindent
which accounts for approximately 5\% of the proton and $N^\ast$ masses.  It is entirely distinct from the HB-only contribution. 

Conversely, and still as a consequence of EHM via its DCSB corollary, the pion and kaon remain massless in the chiral limit. 
As the SM Nambu--Goldstone modes~\cite{Nambu:1960tm, Goldstone:1961eq, Gell-Mann:1968hlm, Horn:2016rip}, their masses vanish due to the precise cancellation between the emergent components of the \( q \) and \( \bar{q} \) masses and the attractive interaction between them; see Eq.\,\eqref{EasyOne}.  

Probing the momentum dependence of the dressed-quark mass, as seen in the structural observables of pions, kaons, and ground and excited state nucleons, allows us to explore EHM in different environments. 
Studies of pion and kaon structure provide valuable insights into the interplay between the Higgs and emergent contributions to hadron mass generation; and given Eq.\,\eqref{GT_eqs}, EHM itself via studies of pseudoscalar meson elastic electromagnetic form factors.
Complementary studies are enabled by investigations of the nucleon elastic form factors and $\gamma_v p N^\ast$ electrocouplings as a function of $Q^2$, which enable one to map the emergent contribution to the dressed-quark and baryon masses and their environmental dependence.  

Consistent results for the dressed-quark mass function from independent studies of nucleon ground and excited states, as well as pion and kaon structure, will reinforce our understanding of the momentum dependence of the dressed-quark mass in a nearly model-independent way. Furthermore, 
they will deepen our insight into EHM and its connection to DCSB.  Connections to confinement can be drawn via fragmentation functions, predictions for which are becoming available~\cite{Xing:2023pms, Xing:2025eip}.

\begin{figure}[t]%
\centering
\includegraphics[width=0.95\textwidth]{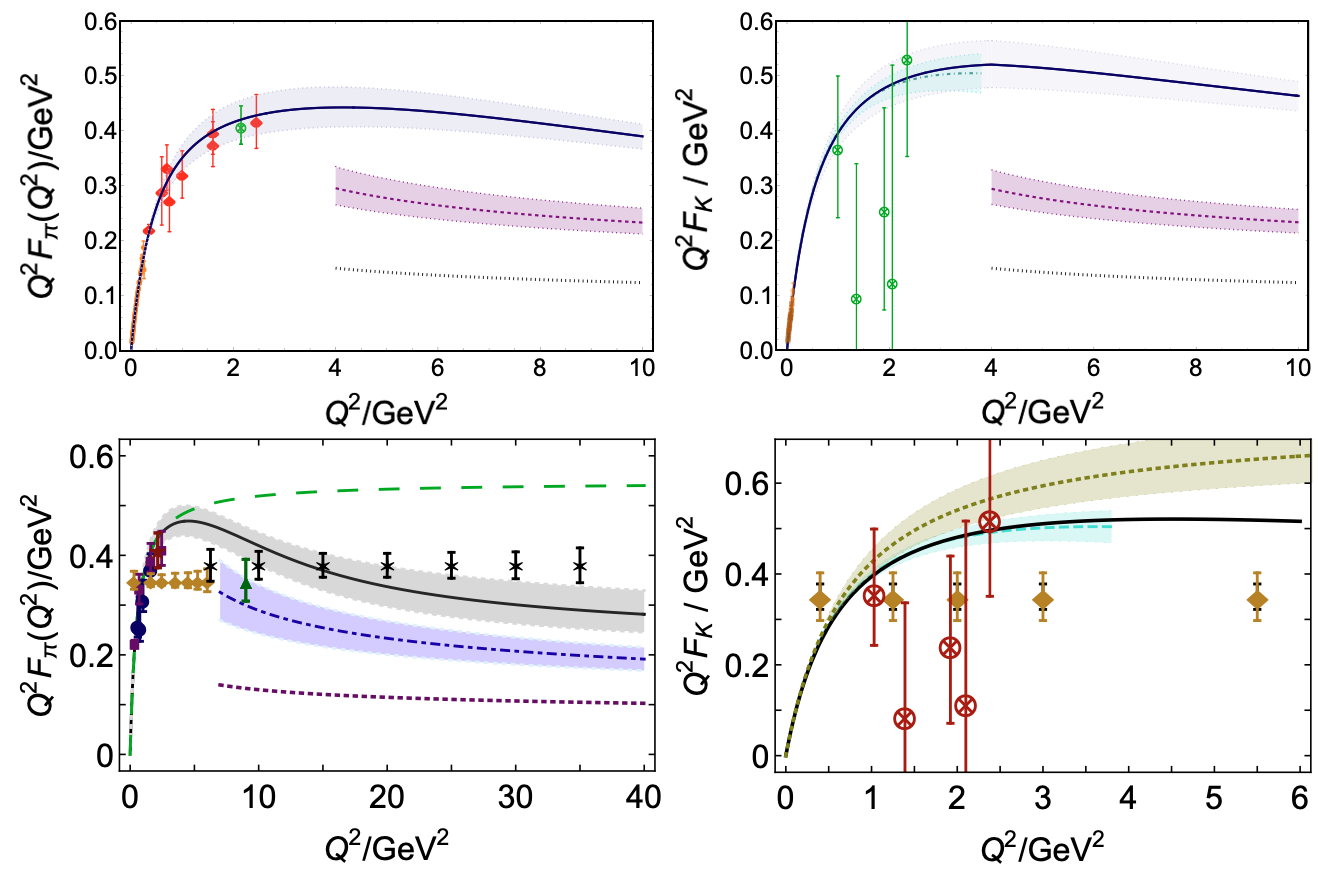}
\caption{(Left, top) CSM predictions for the $\pi^+$~\cite{NA7:1986vav, Horn:2007ug, JeffersonLab:2008jve} and (right, top) $K^+$ elastic electromagnetic form factors~\cite{Dally:1980dj, Amendolia:1986ui, Carmignotto:2018uqj} in comparison with experimental results~\cite{Xu:2023izo}. 
The projected results for 12-GeV era experiments in Halls A/C at JLab for the $\pi^+$ and $K^+$ form factors~\cite{Dudek:2012vr} are shown by the brown diamonds in the bottom left and right graphs, respectively. The black crosses in the bottom left graph represent the projections for the results expected from the Electron--Ion Collider~\cite{Aguilar:2019teb, Arrington:2021biu}.
\label{pipkpff}}
\end{figure}

CSM predictions for the $Q^2$ evolution of the $\pi^+$ and $K^+$ electromagnetic form factors~\cite{Raya:2024ejx} in comparison with the available experimental results are shown in Fig.\,\ref{pipkpff} (top row). 
The CSM predictions were obtained within a unified framework for the description of meson and baryon structure, which exploits insights into QCD's Schwinger functions drawn from studies of QCD's Dyson-Schwinger equations.
An excellent description of all available experimental data for the $\pi^+$ electromagnetic form factor~\cite{NA7:1986vav, Horn:2007ug, JeffersonLab:2008jve} has been achieved, with a $\chi^2/\text{d.p.}=1.0$~\cite{Raya:2024ejx} (see Fig.\,\ref{pipkpff}, top row, left panel). 
However, the large uncertainties in the experimental results for the $K^+$ electromagnetic form factor~\cite{Dally:1980dj, Amendolia:1986ui, Carmignotto:2018uqj, Cui:2021aee} (Fig.~\ref{pipkpff}, top row, right panel) hinder a conclusive assessment of the consistency between these data and CSM predictions.

Studies of Sullivan processes in 12-GeV era experiments in Halls A and C at JLab~\cite{huber-2006,horn-2007} will enable the extension of our understanding of the $Q^2$ evolution of the $\pi^+$ and $K^+$ elastic electromagnetic form factors for $Q^2 < 6$\,GeV$^2$. 
The projected accuracy of these measurements as a function of $Q^2$ is indicated by the brown diamonds in Fig.\,\ref{pipkpff} (bottom row). 
The $\pi^+$ form factor is expected to be determined with an accuracy comparable to the uncertainties of currently available results~\cite{Horn:2007ug, JeffersonLab:2008jve}. 
For the $K^+$ form factor, the anticipated uncertainties will be significantly reduced, enabling a meaningful comparison with CSM predictions. 
The results for the $\pi^+$ elastic form factor from Sullivan process studies at the EIC~\cite{Aguilar:2019teb, Arrington:2021biu} will extend the $Q^2$ coverage up to 35\,GeV$^2$, allowing a detailed exploration of the full range of distances where the emergent part of the dressed-quark mass is generated in the pseudoscalar meson sector. 
Confirming the momentum dependence of the dressed-quark mass through baryon structure studies is essential to establish the relevance of dressed-quarks with dynamically generated masses as an active component in hadron structure.

\begin{figure}[t]
\begin{center}
\includegraphics[width=0.85\textwidth]{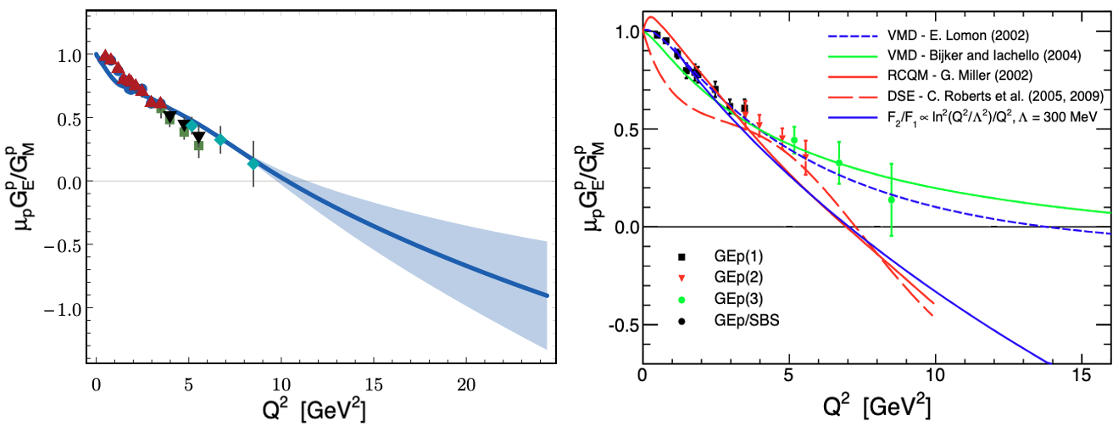}
\end{center}
\caption{Ratios of Sachs form factors, $\mu_p G_E^p(Q^2)/G_M^p(Q^2)$, for the proton. 
Left: CSM prediction in Ref.\,\cite{Cui:2020rmu} compared with data (brown up-triangles~\cite{JeffersonLabHallA:1999epl}; green squares~\cite{JeffersonLabHallA:2001qqe}; blue circles~\cite{Punjabi:2005wq}; black down-triangles~\cite{Puckett:2011xg}; and cyan diamonds~\cite{Puckett:2010ac}). 
Right: Available results for experiments of 12-GeV era in Hall A/C at JLab for $\mu_p G_E^p(Q^2)/G_M^p(Q^2)$ (also referred to as GEp/SBS) from Refs.\,\cite{Barabanov:2020jvn}. 
In the left panel, the $1\sigma$ uncertainty associated with the CSM algorithm for reaching large $Q^2$ is shaded in light blue.}
\label{GEonGM}
\end{figure}

Available experimental results for the ratio of the proton electric to magnetic form factors, $\mu_{p} G_E(Q^2) / G_M(Q^2)$, are shown in Fig.\,\ref{GEonGM} (left) in comparison with the CSM predictions~\cite{Cui:2020rmu}. 
The CSM evaluations of the form factor ratio were obtained with the same momentum dependence of the dressed-quark mass as used in the evaluations of the $\gamma_v p N^\ast$ electrocouplings discussed in Section~\ref{6gev_baryons}. 
A good description of the experimental results has been achieved across the entire $Q^2$ range where measurements are available. 
This success further solidifies the previously established evidence available from the studies of the $\gamma_vpN^\ast$ electrocouplings for insights into the dressed-quark mass function within the range of distances where emergent mass is the dominant determining characteristic of dressed valence quarks with a baryon. 

The 12-GeV era experiments with the Super Bigbite Spectrometer (SBS) at JLab~\cite{jones-2010} will significantly extend the $Q^2$ range of measurements of nucleon elastic form factors. 
The SBS offers a unique opportunity to explore exclusive electroproduction reactions, including cross section measurements and the polarization of reaction products, at the highest luminosities ever achieved — up to $10^{38}$~cm$^{-2}$s$^{-1}$ within a small acceptance (solid angle of msr). These measurements, which are already in progress~\cite{Christy:2021snt}, will provide valuable information on nucleon electromagnetic form factors for $Q^2 < 15$\,GeV$^2$. 
The expected accuracy of these results is shown in Fig.\,\ref{GEonGM}. 

CSMs predicted the presence of a zero crossing~\cite{Cui:2020rmu} in the $Q^2$ evolution of the ratio $\mu_{p} G_E / G_M(Q^2) $ at around 10~GeV$^2$.
That study used the same dressed-quark mass function employed in the evaluations 
of the electrocouplings of the $\Delta(1223)3/2^+$, $N(1440)1/2^+$, and $\Delta(1600)3/2^+$. 
Such CSM calculations have demonstrated that the location of the zero at a specific $Q^2$ is related to the rate of sQCD\,$\leftrightarrow$\,pQCD transition, as reflected in the momentum dependence of the dressed-quark mass~\cite{Wilson:2011aa, Cloet:2013gva}.  
Contemporary CSM analyses, which deliver a single-framework unification of  pseudoscalar meson and nucleon properties, have confirmed the earlier results~\cite{Yao:2024drm, Yao:2024uej}.

Recent extrapolations of experimental data on the $\mu_{p} G_E / G_M(Q^2)$ ratio, performed using the Slessinger Point Method (SPM)~\cite{Cheng:2024cxk}, suggest that, with 50\% confidence, the zero crossing is located at $Q^2 < 10.4$~GeV$^2$. 
The confidence level increases to 99.9\% for $Q^2 < 13.1$~GeV$^2$. Alternatively, the likelihood that the existing data are consistent with the absence of a zero within $Q^2 < 14.5$~GeV$^2$ is only $10^{-6}$.  

The range $Q^2 < 15$\,GeV$^2$ accessible in the 12-GeV era experiments will enable the localization of the zero of the $\mu_{p} G_E / G_M(Q^2)$ ratio in the $Q^2$ evolution and allow the determination of the transition rate from the sQCD to the pQCD regime directly from the experimental data. Confirming this transition rate through independent studies of the pion and kaon electromagnetic form factors, as well as the $\gamma_v p N^\ast$ electrocouplings of different nucleon resonances, is crucial to strengthen the evidence for insights into the momentum dependence of the dressed-quark mass.


\section{Conclusions and Outlook}
Nucleons and their excited state resonances represent the most fundamental three-body systems in Nature. 
If we do not fully understand how the prominent states in hadron spectra emerge from QCD, then our comprehension of strong interaction dynamics and its evolution with distance 
is incomplete. 
Studies of both meson and baryon structure address key unresolved problems in the SM, including the emergence of hadron mass and the origin of hadron structure in relation to the roles of DCSB and quark and gluon confinement 
\cite{Roberts:2015lja, Horn:2016rip, Roberts:2021nhw, Binosi:2022djx, Ding:2022ows, Carman:2023zke, Ferreira:2023fva, Raya:2024ejx}.

The concept of EHM, developed using CSMs, currently provides the only available approach that enables insight into the dynamics of hadron mass generation from experimental data on hadron structure observables. 
It does so within a unified theoretical framework that is applicable to both mesons and baryons.
Extending this framework through complementary approaches that connect to the QCD Lagrangian --- as, \textit{e.g}., with lQCD --- is highly desirable, so that we may deepen our understanding of hadron mass and structure.

High-quality meson electroproduction data from the 6-GeV era with CLAS have enabled the determination of the electrocouplings for most $N^\ast$ states in the mass range up to 1.8\,GeV for photon virtualities $Q^2 < 5$\,GeV$^2$~\cite{Carman:2023zke}. 
In the coming years, the electrocouplings of the most prominent resonances up to 2\,GeV are expected to become available for $Q^2 < 5$\,GeV$^2$.

A good description of the $\Delta(1232)3/2^+$ and $N(1440)1/2^+$ electrocouplings for $Q^2 < 5$\,GeV$^2$, achieved using CSMs in studies that employ a common dressed quark mass function, developed via analyses of the QCD Lagrangian, validated by the successful descriptions of pion and nucleon elastic electromagnetic form factors, provides strong evidence of meaningful insight into the momentum dependence of the dressed quark mass. 
Moreover, the 2019 CSM prediction for the $Q^2$ evolution of the $\Delta(1600)3/2^+$ electrocouplings has been confirmed by the first experimental results obtained in 2023~\cite{Mokeev:2023zhq}, further solidifying evidence for insight into EHM through the study of the $\gamma_v p N^\ast$ electrocouplings. 
Preliminary empirical results on the $\Delta(1700)3/2^-$ electrocouplings have also recently become available~\cite{mokeev-nstar24}. 
Comparative studies of the electrocouplings of the chiral partners $\Delta(1232)3/2^+$ and $\Delta(1700)3/2^-$ now offer a promising opportunity to explore EHM in connection with DCSB.

Global CSM analyses of the $\gamma_v p N^\ast$ electrocouplings for $N^\ast$ states with masses up to 2\,GeV and photon virtualities $Q^2 < 5$\,GeV$^2$, along with studies of the structure of the ground-state nucleon, pion, and kaon, are essential for understanding EHM~\cite{Ding:2022ows, Horn:2016rip, Carman:2023zke, Raya:2024ejx}. 
Investigations of the structure of both the ground and excited nucleon states, along with the structure of pseudoscalar mesons, provide complementary perspectives on hadron mass generation. 
The elastic form factors and parton distribution functions of pseudoscalar mesons are particularly sensitive to the interplay between the Higgs mechanism and emergent mass generation dynamics. 
Likewise, with strong complementarity, studies of ground and excited state nucleon structure probe the emergent component of hadron mass. 
Comprehensive analyses that combine experimental results on pseudoscalar meson structure from Sullivan processes at JLab and Drell-Yan processes at COMPASS and AMBER@CERN, with the extensive data on nucleon structure already available --- and anticipated from ongoing 12-GeV experiments at JLab --- are crucial for advancing our understanding of EHM.

The kinematic coverage of the 6-GeV era experiments at JLab enables exploration of the distance (momentum) range where up to 30\% of the dressed-quark mass is expected to be contained. 
With the ongoing experiments of the 12-GeV era at JLab, CLAS12 is currently the only facility in the world capable of extracting the electrocouplings of all prominent $N^\ast$ states in the still unexplored $Q^2$ range from 5 to 10\,GeV$^2$, through measurements of the $\pi N$, $\pi^+\pi^-p$, $\eta N$, $K\Lambda$, and $K\Sigma$ exclusive electroproduction channels. 
These measurements will allow for detailed mapping of the dressed quark mass function at quark momenta corresponding to the region where approximately 50\% of the emergent mass is localized.

Following a possible energy upgrade of CEBAF to 22\,GeV, and with the capability to perform exclusive meson electroproduction measurements at luminosities in the range of $5 \times 10^{35}$\,cm$^{-2}$s$^{-1}$, we will be able to map the full range of distances over which almost 100\% of mass and structure emerge~\cite{Accardi:2023chb}. 
In particular, this upgrade will make it possible to extend experimental access to the $Q^2$ evolution of the $\gamma_v p N^\ast$ electrocouplings for prominent nucleon resonances, covering a $Q^2$ range up to 30\,GeV$^2$. 
These measurements will allow for detailed insight into the evolution of the dressed quark mass over the complete distance scale where the emergent component of hadron mass is generated and the transition to pQCD has been completed.

It is worth stressing that such studies will not be feasible at any other facility foreseen in the world --- including the electron--ion colliders (EIC in the U.S.\ and EicC in China) --- owing to limitations of more than an order of magnitude lower luminosity than needed to explore nucleon resonance electroexcitations for $Q^2 < 30$\,GeV$^2$ and restricted coverage in the accessible invariant mass $W$ of the final hadronic system, which is far above the resonance excitation region; see Fig.\,\ref{facilities}. 
From this perspective, a CEBAF energy upgrade to 22\,GeV could be seen as delivering the ultimate QCD facility at the luminosity frontier.

Forging synergistic efforts between experiment, phenomenology, and QCD-based hadron structure theory is essential for addressing the key open problems in the SM that were outlined above. 
The success of the CSM framework in advancing our understanding of hadron mass and structure generation both invites and paves the way for broader theoretical efforts. 
These efforts will be critical in supporting the interpretation of experimental results on hadron structure, helping to resolve long-standing challenges within the SM, and potentially offering insights into the deeper complexity of Nature that lies beyond its current boundaries.

\vspace{6pt} 



\funding{
This material is based upon work supported in part by 
U.S.\ Department of Energy, Office of Science, Office of Nuclear Physics under contract DE-AC05-06OR23177 (PA, DSC, VIM); 
National Science Foundation under Grant PHY 10014377 (RWG); 
and
National Natural Science Foundation of China under grant no.\ 12135007 (CDR).
}

\dataavailability{Data sharing is not applicable.} 

\acknowledgments{This contribution is based on results obtained and insights developed through collaborations with many people, to all of whom we are greatly indebted.
}

\conflictsofinterest{The authors declare no conflicts of interest.} 


\abbreviations{Abbreviations}{
The following abbreviations are used in this manuscript:\\

\noindent 
\begin{tabular}{@{}ll}
CEBAF & Continuous Electron Beam Accelerator Facility at Jefferson Lab\\
CERN & European Organization for Nuclear Research\\
CLAS & CEBAF Large Acceptance Spectrometer\\
CLAS12 & CEBAF Large Acceptance Spectrometer for use at 12 GeV\\
CM & Center-of-mass frame\\
CSM(s) & Continuum Schwinger function method(s)\\
DCSB & Dynamical Chiral Symmetry Breaking\\
d.p.\ & Data point\\
EHM & Emergence of hadron mass\\
EIcC & Electron Ion Collider in China\\
EIC & Electron Ion Collider at Brookhaven National Laboratory\\
EMT & Energy-Momentum Tensor\\
HB & Higgs boson\\
JLab & Thomas Jefferson National Accelerator Facility -- Jefferson Laboratory\\
JM & JLab-Moscow data-driven reaction model\\
lQCD & Lattice Quantum Chromodynamics\\
PDG & Particle Data Group\\
PI (charge) & Process-independent (charge) \\
pQCD & Perturbative Quantum Chromodynamics\\
QCD & Quantum Chromodynamics\\
QED & Quantum Electrodynamics\\
QFT & Quantum Field Theory\\
RGI & Renormalization Group Invariant\\
RL & Rainbow Ladder\\
SAID & Partial Wave Analysis Facility at George Washington University\\
SBS & Super Bigbite Spectrometer at Jefferson Lab\\
SCI & Symmetry preserving contact interaction\\
SM & Standard Model of nuclear and particle physics\\
SPM & Schlessinger Point Method\\
sQCD & Strongly coupled QCD
\end{tabular}
}

\begin{adjustwidth}{-\extralength}{0cm}

\reftitle{References}
\bibliography{References}

%

\PublishersNote{}
\end{adjustwidth}
\end{document}